\documentclass[12pt]{article}
\usepackage{epsfig}
\usepackage{lineno}
\usepackage{amsmath}
\usepackage{hhline}
\usepackage{amssymb}
\usepackage{times}
\usepackage{cite}

\newlength{\dinwidth}
\newlength{\dinmargin}
\newcommand{\xgobs}{$x_{\gamma}^{\textrm{obs}}$}
\newcommand{\xgobsm}{x_{\gamma}^{\textrm{obs}}}
\newcommand{\ptjet}{$p_{\textrm{t}}^{\textrm{\,jet}_{1}}$}
\newcommand{\ptjetm}{p_{\textrm{t}}^{\textrm{\,jet}_{1}}}
\newcommand{\ptjettwo}{$p_{\textrm{t}}^{\textrm{\,jet}_{2}}$}

\newcommand{\ptjetsm}{p_{\textrm{t}}^{\textrm{\,jet}_{1(2)}}}
\newcommand{\pt}{$p_{\textrm{t}}$}
\newcommand{\ptm}{p_{\textrm{t}}}
\newcommand{\fbbb}{$f^{\textrm{b}\bar{\textrm{b}}}$}
\newcommand{\fccb}{$f^{\textrm{c}\bar{\textrm{c}}}$}

\begin{document}

\pagestyle{empty}

\begin{titlepage}

\noindent
\begin{flushleft}
DESY 06-039 \\
April 2006
\end{flushleft}

\vspace*{2cm}

\begin{center}
  \Large
  {\bf 
      Measurement of Charm and Beauty\\
      Dijet Cross Sections
      in Photoproduction at HERA \\
      using the H1 Vertex Detector
      }

  \vspace*{2cm}
    {\Large H1 Collaboration} 
\end{center}


\begin{abstract}

\noindent
A measurement of charm and beauty dijet photoproduction 
cross sections at the $ep$ collider HERA is presented.
Events are selected with two or more jets of transverse momentum
$p_{\textrm{t}}^{\textrm{jet}_{1(2)}}>11(8)$ GeV in the central range 
of pseudo-rapidity $-0.9<\eta^{\textrm{jet}_{1(2)}}<1.3$.
The fractions of events containing charm and beauty quarks 
are determined using a method based on the impact parameter, in the 
transverse plane, of tracks to the primary vertex, as measured by 
the H1 central vertex detector.
Differential dijet cross sections for charm and beauty, and their
relative contributions to the flavour inclusive dijet photoproduction 
cross section, are measured as a function of the transverse momentum 
of the leading jet, the average pseudo-rapidity of the two jets
and the observable $\xgobsm$. 
Taking into account the theoretical uncertainties, the charm cross 
sections are consistent with a QCD calculation in next-to-leading
order, while the predicted cross sections for beauty production are
somewhat lower than the measurement. 

\end{abstract}

\vspace{1cm}
\centerline{Submitted to Eur.\,Phys.\,J. C}

\end{titlepage}
\noindent

\noindent
A.~Aktas$^{9}$,                
V.~Andreev$^{25}$,             
T.~Anthonis$^{3}$,             
B.~Antunovic$^{26}$,           
S.~Aplin$^{9}$,                
A.~Asmone$^{33}$,              
A.~Astvatsatourov$^{3}$,       
A.~Babaev$^{24, \dagger}$,     
S.~Backovic$^{30}$,            
A.~Baghdasaryan$^{37}$
% ...

\bigskip{\it
\noindent
% ...
$ ^{3}$ Inter-University Institute for High Energies ULB-VUB, Brussels;
          Universiteit Antwerpen, Antwerpen; Belgium$^{ c}$ \\
% ...
$ ^{9}$ DESY, Hamburg, Germany \\
% ...
$ ^{24}$ Institute for Theoretical and Experimental Physics,
          Moscow, Russia$^{ k}$ \\
 $ ^{25}$ Lebedev Physical Institute, Moscow, Russia$^{ e}$ \\
 $ ^{26}$ Max-Planck-Institut f\"ur Physik, M\"unchen, Germany \\
% ...
$ ^{30}$ Faculty of Science, University of Montenegro,
          Podgorica, Serbia and Montenegro$^{ e}$ \\
% ...
$ ^{33}$ Dipartimento di Fisica Universit\`a di Roma Tre
          and INFN Roma~3, Roma, Italy \\
% ...
$ ^{37}$ Yerevan Physics Institute, Yerevan, Armenia \\
% ...
$ ^{\dagger}$ Deceased \\

\bigskip
\noindent
% ...
$ ^c$ Supported by FNRS-FWO-Vlaanderen, IISN-IIKW and IWT
      and  by Interuniversity
Attraction Poles Programme,
      Belgian Science Policy \\
% ...
$ ^e$ Supported by the Deutsche Forschungsgemeinschaft \\
% ...
$ ^k$ Partially Supported by Russian Foundation
      for Basic Research,  grants  03-02-17291
      and  04-02-16445 \\
}

\newpage


\pagestyle{plain}
%
\section{Introduction}

A measurement is presented of charm and beauty production in 
$ep$ collisions at HERA using events with
two or more jets at high transverse momentum.
The measurement is carried out in the photoproduction region 
in which a quasi-real photon, with virtuality $Q^2 \sim 0$,
is emitted from the incoming positron and interacts
with a parton from the proton.
Differential charm and beauty dijet cross sections 
are measured and compared to calculations in perturbative quantum 
chromodynamics (pQCD) performed to next-to-leading order (NLO).

In pQCD calculations, the photoproduction of charm and beauty
proceeds dominantly via the direct photon-gluon fusion process
$\gamma g \rightarrow c\bar{c}$ or $b\bar{b}$, where the photon interacts 
with a gluon from the proton to produce a pair of heavy quarks in the 
final state. Previous charm measurements have confirmed this 
prediction~\cite{Aid:1996hj,Breitweg:1998yt,Chekanov:2003bu,Chekanov:2005zg}.  
In leading order QCD models a successful description of the data is 
obtained when additional contributions from processes involving 
resolved photons are taken into account~\cite{Breitweg:1998yt,Chekanov:2003bu}. 
In such resolved photon processes the quasi-real photon fluctuates into a 
hadronic state before the hard interaction and thus acts as 
a source of partons. In the massless scheme, a large fraction of these 
resolved photon processes is due to heavy quark excitation, in which one 
of the partons that enters the hard interaction is a heavy quark 
($c$ or $b$) originating from the resolved photon or the proton. 

In this analysis events containing heavy quarks are distinguished 
from light quark events by the long lifetimes of $c$ and $b$ flavoured hadrons, 
which lead to displacements of tracks from the primary vertex.
This technique, based on the precise spatial information 
from the H1 silicon vertex detector, was introduced in 
recent H1 measurements of the charm and beauty structure 
functions $F_{2}^{\textrm{c}\bar{\textrm{c}}}$ and 
$F_{2}^{\textrm{b}\bar{\textrm{b}}}$ 
in deep inelastic scattering~\cite{Aktas:2004az,Aktas:2005iw},
and is now applied to dijet events in photoproduction.
This analysis provides the first simultaneous measurement 
of charm and beauty in photoproduction, extending to larger values 
of transverse jet momentum than previous 
measurements~\cite{Aid:1996hj,Breitweg:1998yt,Chekanov:2003bu,Chekanov:2005zg,
Adloff:1999nr,Breitweg:2000nz,Chekanov:2004xy,Aktas:2005zc}.
The regions of small transverse momentum and large pseudo-rapidities 
are excluded from the measurement, due to trigger requirements
and the limited angular acceptance of the vertex detector.

The differential charm and beauty dijet cross sections are measured as 
functions of the transverse momentum of the leading jet \ptjet, 
of the mean pseudo-rapidity $\bar{\eta}$ of the two jets, and of the 
variable $\xgobsm$ which, in a leading order picture,
corresponds to the fraction of the photon's energy in the proton rest 
frame that enters the hard interaction. 
For direct photon-gluon fusion processes $\xgobsm \sim 1$, while
for resolved photon processes $\xgobsm$ can be small.
The measured differential charm and beauty cross sections, together
with the measured flavour inclusive cross sections, are used 
to determine the relative contribution from charm
and beauty events to dijet photoproduction.
The results are compared with calculations in perturbative quantum 
chromodynamics at next-to-leading order and with predictions from 
Monte Carlo simulations in which leading order matrix elements are 
implemented, and contributions from higher orders are approximated 
using parton showers.

This paper is structured as follows: 
In section~\ref{sec:h1} the experimental apparatus is briefly described.
Event and track selections are detailed in section~\ref{sec:sel}.
The method to determine the contributions of charm and beauty events 
is outlined in section~\ref{sec:impactparm}.
Theoretical calculations performed in the framework of perturbative 
QCD are discussed in section~\ref{sec:nlo}.
The cross section measurements and their systematic uncertainties 
are presented in section~\ref{sec:crosssections}. 
Properties of a heavy quark enriched data sample
are investigated in section~\ref{sec:crosscheck}.
A summary of the results is given in section~\ref{sec:conclusions}.

\section{H1 Detector}
\label{sec:h1}

The H1 detector is described in detail in~\cite{Abt:1997xv}.
Charged particles emerging from the $ep$ interaction region
are measured by the central tracking detector (CTD) in the pseudo-rapidity
range $-1.74 < \eta < 1.74$\footnote{The pseudo-rapidity is given by
$\eta = -\ln \; \tan (\theta/2)$, where $\theta$ is
measured with respect to the $z$-axis given by the proton beam direction.}.
The CTD consists of two large cylindrical central jet drift chambers (CJCs), 
two $z$ chambers and two multi-wire proportional chambers arranged concentrically 
around the beam-line in a magnetic field of 1.15 T. 
The CTD provides triggering information based on track segments from the CJC
in the $r$-$\phi$ plane, transverse to the beam direction, and on the $z$ 
position of the vertex from the multi-wire proportional chambers.
The CJC tracks are linked with hits in the Central Silicon Tracking detector 
(CST)~\cite{cst}, which consists of two cylindrical layers of silicon 
strip sensors, surrounding the beam pipe at radii of \mbox{$57.5$ mm} and 
\mbox{$97$ mm} from the beam axis. 
The detector provides hit resolutions of 12 $\mu$m in $r$-$\phi$ 
with an average efficiency of $97\%$.

For CTD tracks with CST hits in both layers the impact parameter $\delta$, 
i.e.\,the transverse distance of closest approach to the nominal 
vertex, can be measured 
with a resolution of $\sigma_{\delta} \approx 33\;\mu\mbox{m} 
\oplus 90 \;\mu\mbox{m}/\ptm [\mbox{GeV}]$.
The first term represents the intrinsic resolution
and includes the uncertainty of the CST alignment,
the second term corresponds to the contribution from multiple scattering 
in the beam pipe and the CST, which depends on the transverse
momentum \pt \ of the track.

Charged and neutral particles are measured in the liquid argon (LAr) 
calorimeter which surrounds the 
tracking chambers and covers the range $-1.5 < \eta < 3.4$ and
a lead--scintillating fibre calorimeter SpaCal, covering the backward 
region ($-4.0 < \eta < -1.4$)~\cite{Nicholls:1996di}.
The measurements from CTD and calorimeters are combined to 
reconstruct the final state particles~\cite{peez}.
The luminosity determination is based on the measurement of the 
Bethe-Heitler process $(ep \to ep\gamma)$, where the photon is detected 
in a calorimeter located downstream of the interaction 
point in the positron beam direction. 

\section{Event and Track Selection}
\label{sec:sel}

The data sample corresponds to an integrated 
luminosity of $56.8\,$pb$^{-1}$ and 
was recorded with the H1 experiment during the years 1999 and 2000. 
During this time HERA was operated with positrons
of 27.5 GeV energy and protons of 920~GeV.
The events were triggered by a combination of signals from the calorimeters, 
the central drift chambers and the multi-wire proportional chambers. 
Photoproduction events are selected by requiring that there be no 
isolated high energy electromagnetic cluster detected in the calorimeters
consistent with a signal from the scattered positron.
This restricts the range of negative four-momentum transfer squared 
to $Q^2<1$ GeV$^2$. 
The inelasticity $y$ is calculated using the hadronic final state~\cite{JB}, 
and the measurement is restricted to the range \mbox{0.15 $<y<$ 0.8}.
The jets are reconstructed from the final state particles
using the inclusive $k_{\textrm{t}}$ algorithm~\cite{kt} in 
the \pt \ recombination scheme~\cite{pt}, with distance parameter 
$R=1$ in the \mbox{$\eta$-$\phi$} plane. 
The event selection requires at least two jets in the central 
pseudo-rapidity range $-0.9 < \eta < 1.3 $ with transverse energy 
\mbox{$\ptjetsm>11(8)$ GeV}. 

CTD tracks are selected which are linked to hits in 
both $r$-$\phi$ layers of the CST. These tracks are required to have 
a transverse momentum above 500 MeV and a polar angle in the range 
$30^\circ<\theta_{\textrm{track}}<150^\circ$. 
Events are selected which contain at least one selected track associated 
to one of the two leading jets. The final sample consists of 80769 events.

To correct for detector effects, such as detector resolutions 
and inefficiencies, large samples of charm, beauty and 
light quark events are generated using the Monte Carlo 
program PYTHIA~\cite{PYTHIA} (for details see section~\ref{sec:nlo}). 
All samples are passed through a detailed simulation 
of the H1 detector response based on the GEANT program~\cite{geant} and the
same reconstruction and analysis algorithms as used for the data.

\begin{figure}[p]
\unitlength1cm
\begin{picture}(8,18)
\put(0.,12.){\includegraphics*[width=7.9cm]{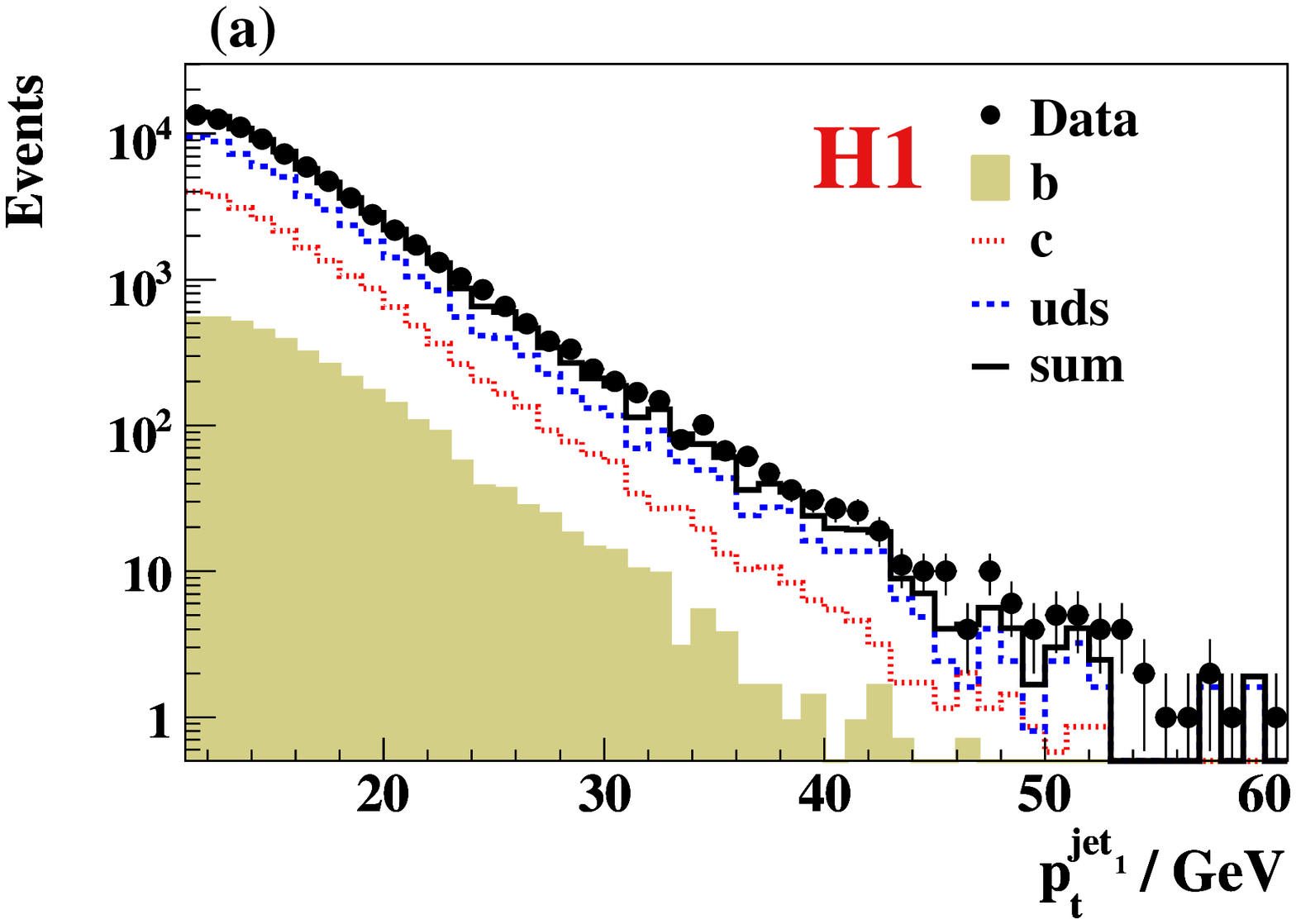}}
\put(8.1,12.){\includegraphics*[width=7.9cm]{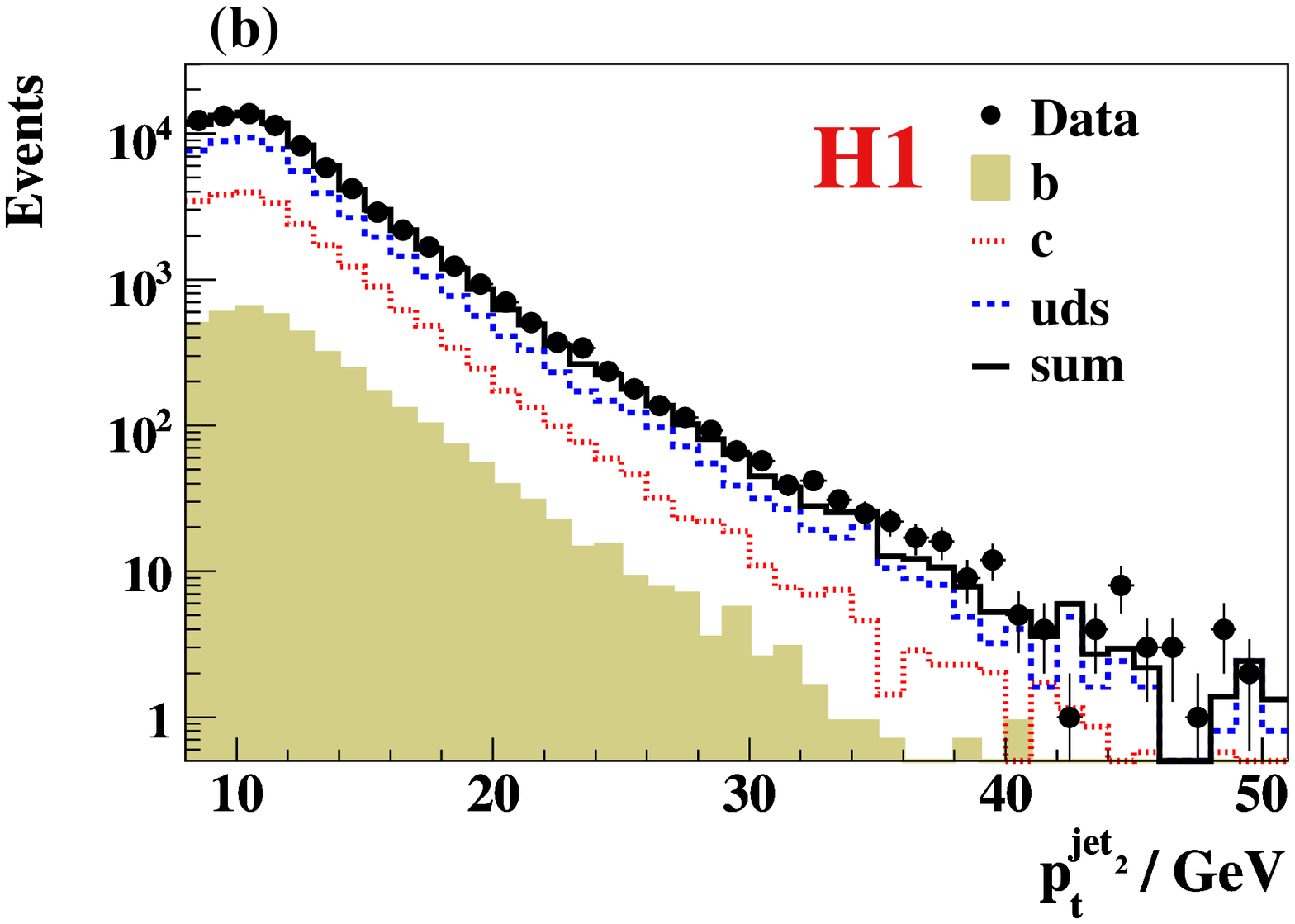}}
\put(0.,6.){\includegraphics*[width=7.9cm]{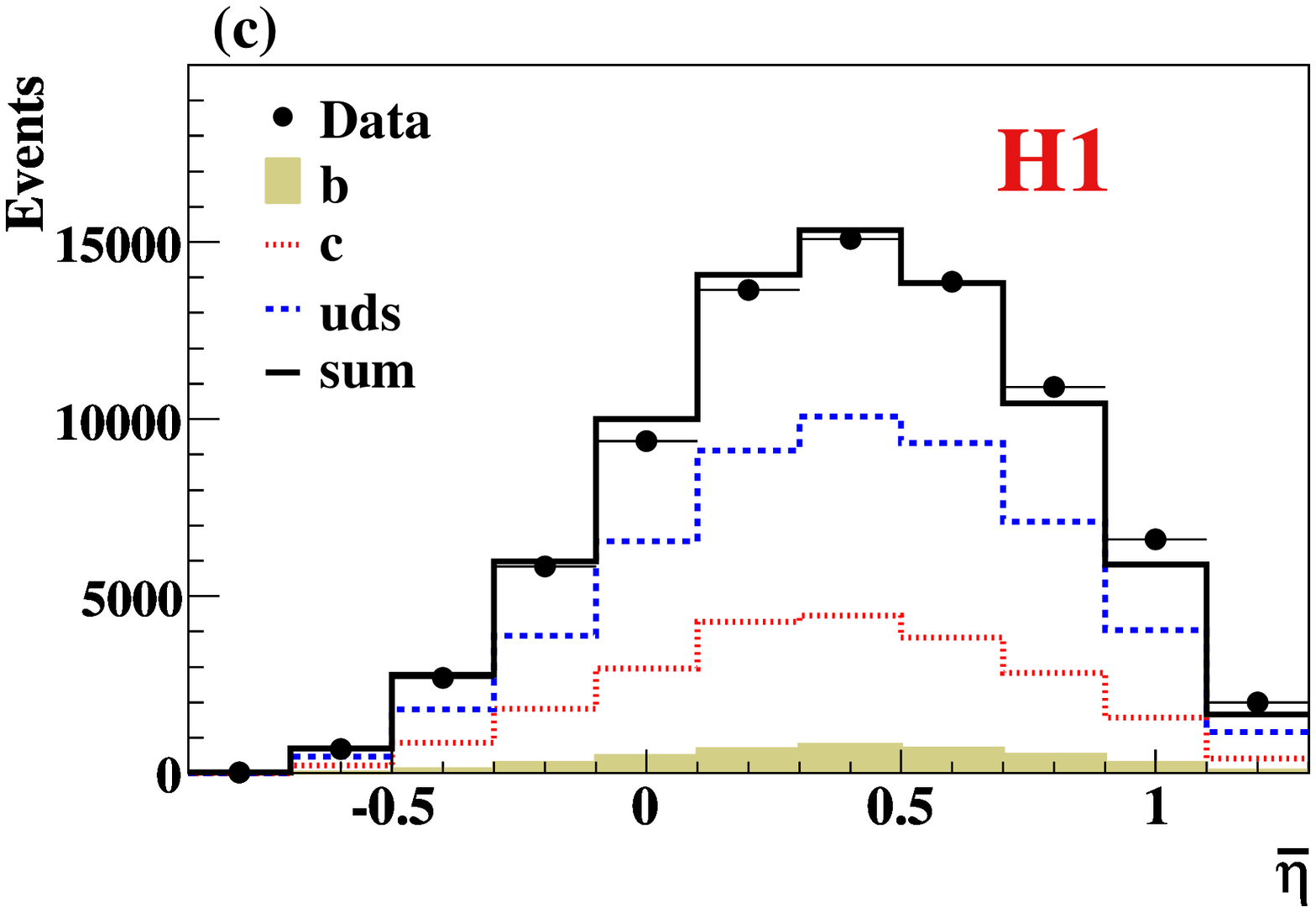}}
\put(8.1,6.){\includegraphics*[width=7.9cm]{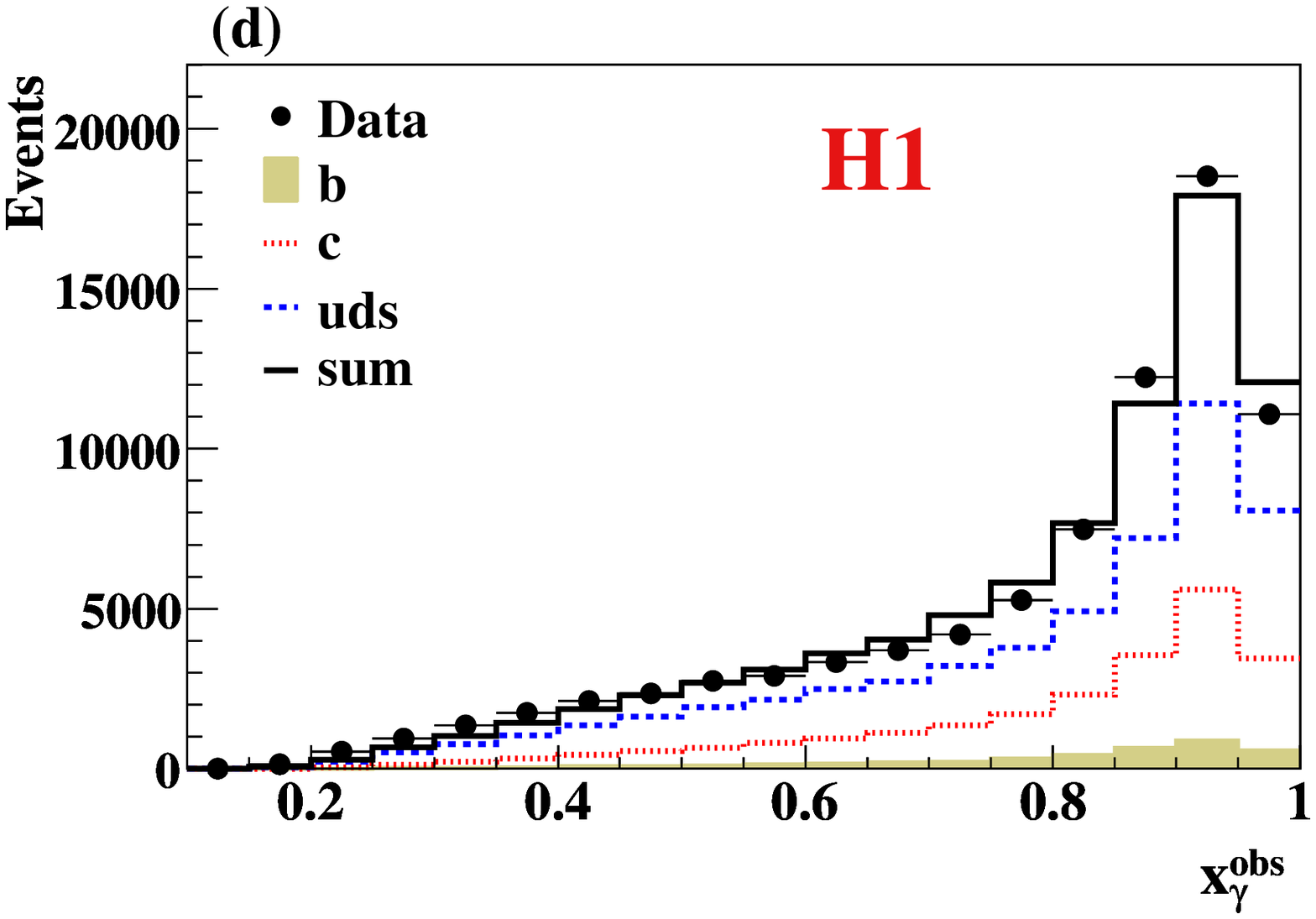}}
\put(0.,0.){\includegraphics*[width=7.9cm]{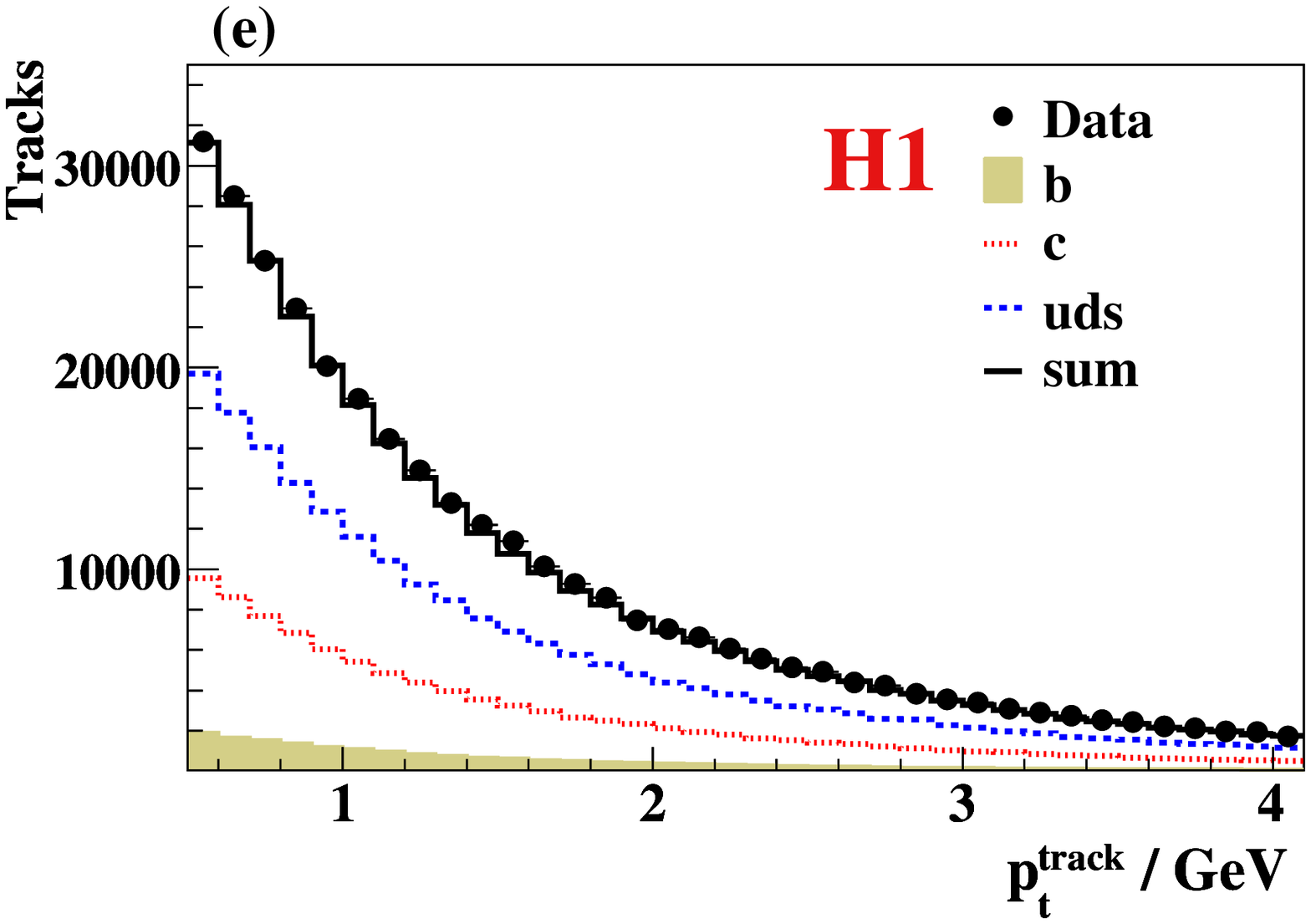}}
\put(8.1,0.){\includegraphics*[width=7.9cm]{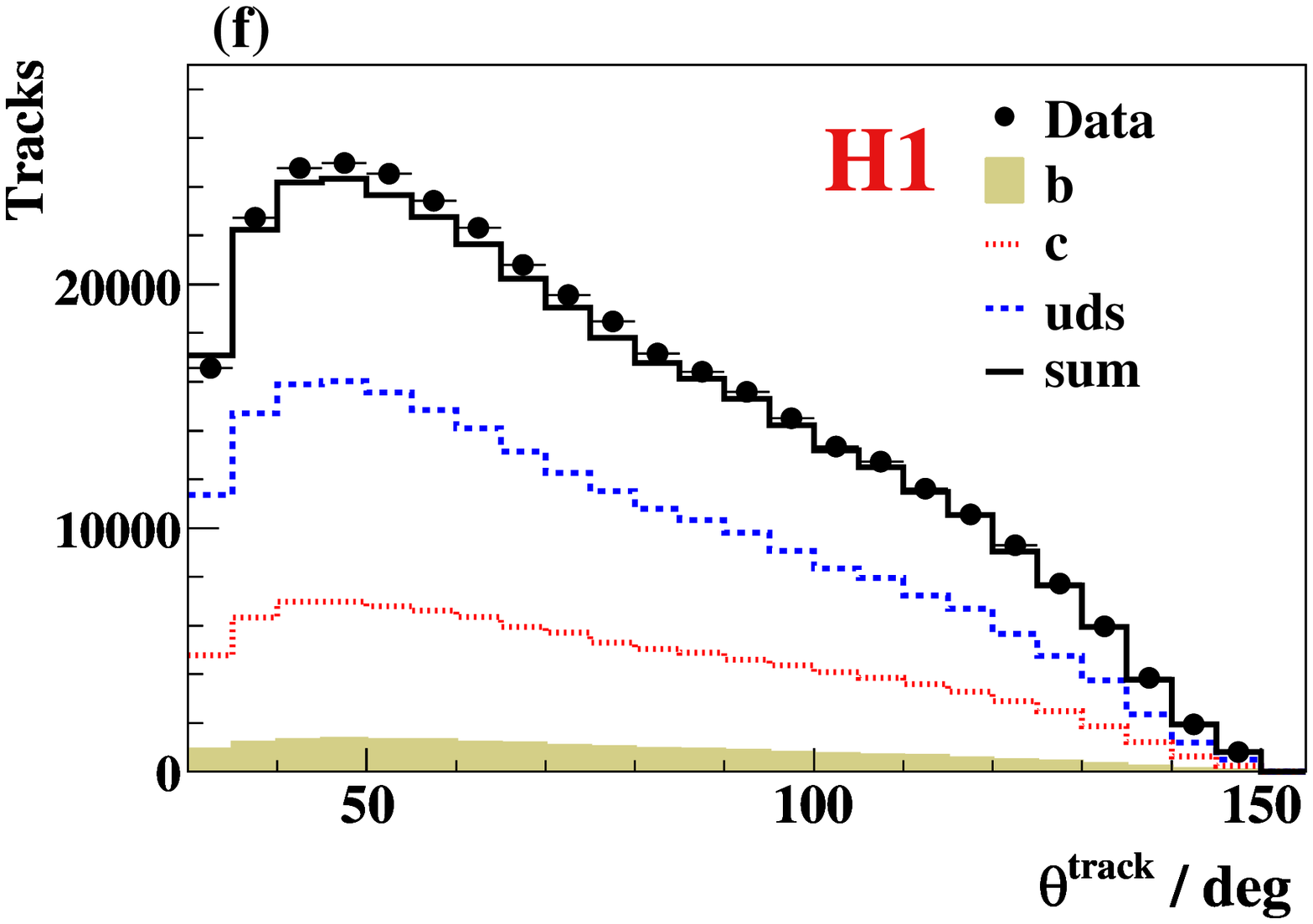}}
\end{picture}
\caption{Distributions of 
  a) the transverse momentum \ptjet \ of the leading jet,
  b) the transverse momentum \ptjettwo \ of the second jet,
  c) the mean pseudo-rapidity $\bar{\eta}$ of the two jets,
  d) the observable $\xgobsm$,
  e) the transverse momentum of the selected tracks and 
  f) the polar angle of the selected tracks.
  The expectation from the PYTHIA
  Monte Carlo simulation is included in the figure, 
  showing the contributions from the various quark flavours after applying 
  the scale factors obtained from the 
  fit to the subtracted significance distributions of the data
  (see section~\ref{sec:impactparm}).}
\label{fig:ctrl:jet1} 
\end{figure}

Figure\,\ref{fig:ctrl:jet1} shows the distributions of the transverse momentum
of the leading jet \ptjet, 
and of 
the jet with the second highest transverse momentum \ptjettwo \ 
(figures~\ref{fig:ctrl:jet1}a and b), of
the mean pseudo-rapidity 
$\bar{\eta}=(\eta^{\textrm{jet}_{1}}+\eta^{\textrm{jet}_{2}})/2$ 
of the two jets (figure~\ref{fig:ctrl:jet1}c)
and of $\xgobsm$ (figure~\ref{fig:ctrl:jet1}d).
The observable $\xgobsm$ is defined as $ ((E-p_{\textrm{z}})_{\textrm{jet}_{1}}
+(E-p_{\textrm{z}})_{\textrm{jet}_{2}})/\sum(E-p_{\textrm{z}})$ where the sum runs 
over all measured particles of the final state. 
In figures~\ref{fig:ctrl:jet1}e and f the transverse momentum and polar 
angular distributions of the selected tracks are shown. 
The simulation based on the PYTHIA event generator provides a
good description of all distributions, after scaling the contributions 
from light quark, charm and beauty events. The scale factors 
are obtained from fits to the signed impact parameter distributions, 
as described below in section~\ref{sec:impactparm}.

\section{Quark Flavour Separation}
\label{sec:impactparm}

The fractions of events containing charm and beauty quarks 
are determined using the same method as in previous 
measurements~\cite{Aktas:2004az,Aktas:2005iw}, based on the
impact parameter of selected tracks which is given by 
the transverse distance of closest approach to the reconstructed event vertex.
The signed impact parameter $\delta$ is defined as positive if the 
angle between the axis of the associated jet and the line joining the 
primary vertex to the point of closest approach of the track 
is less than $90^\circ$, and is defined as negative otherwise.

\begin{figure}[htb]
\unitlength1cm
\begin{picture}(8,18)
\put(0.,12.){\includegraphics*[width=7.9cm]{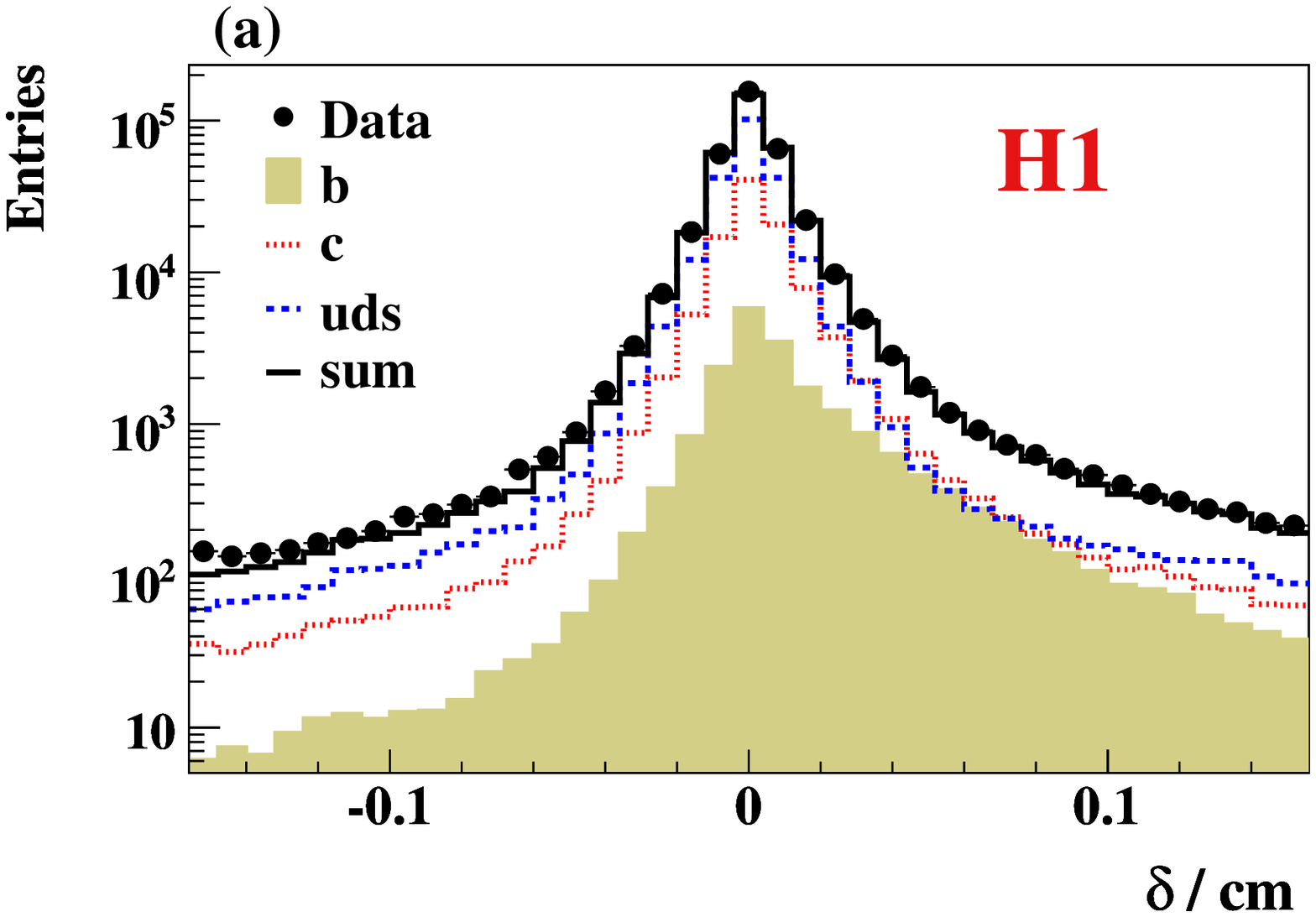}}
\put(8.1,11.9){\includegraphics*[width=7.9cm]{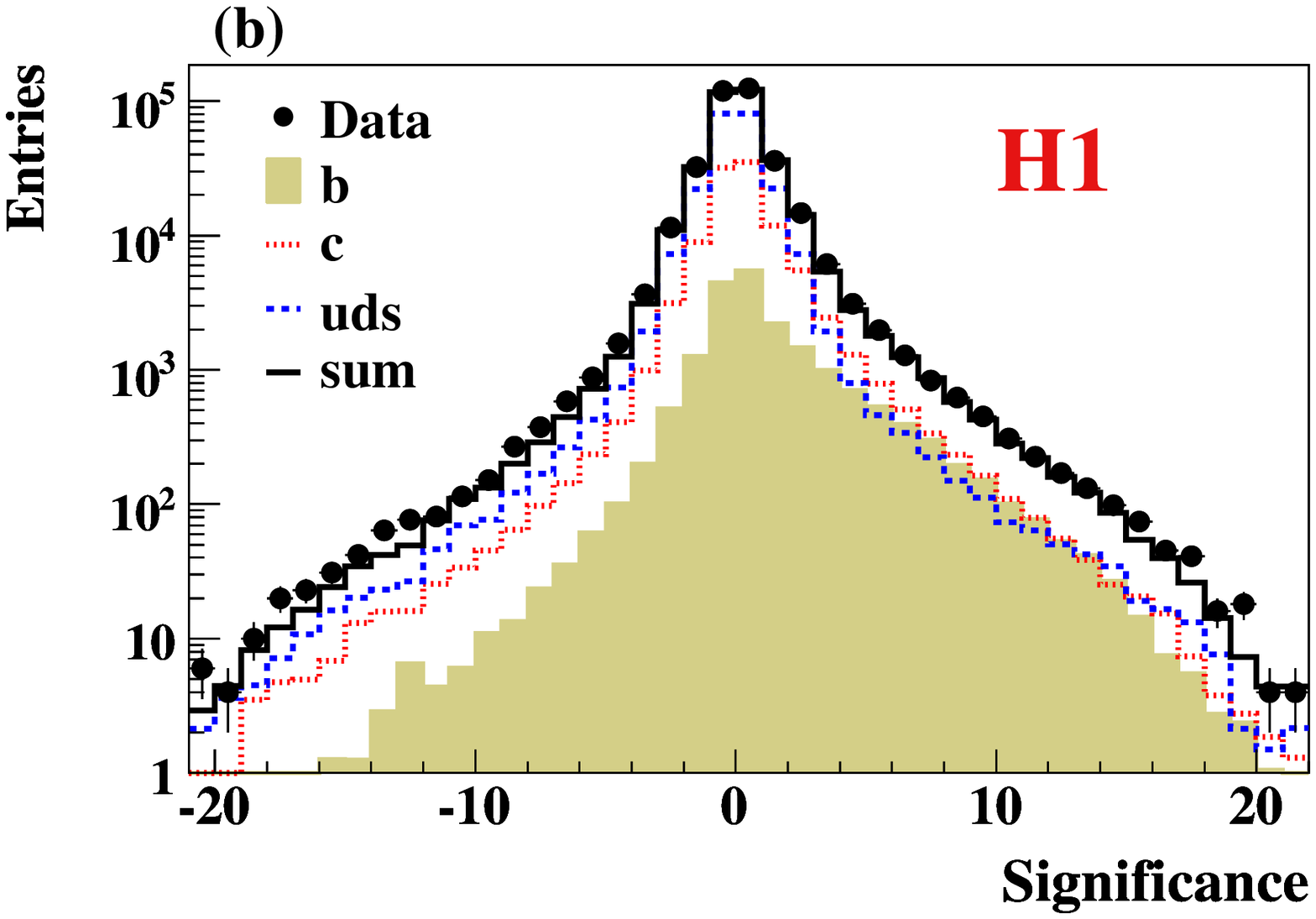}}
\put(0.,6.) {\includegraphics*[width=7.9cm]{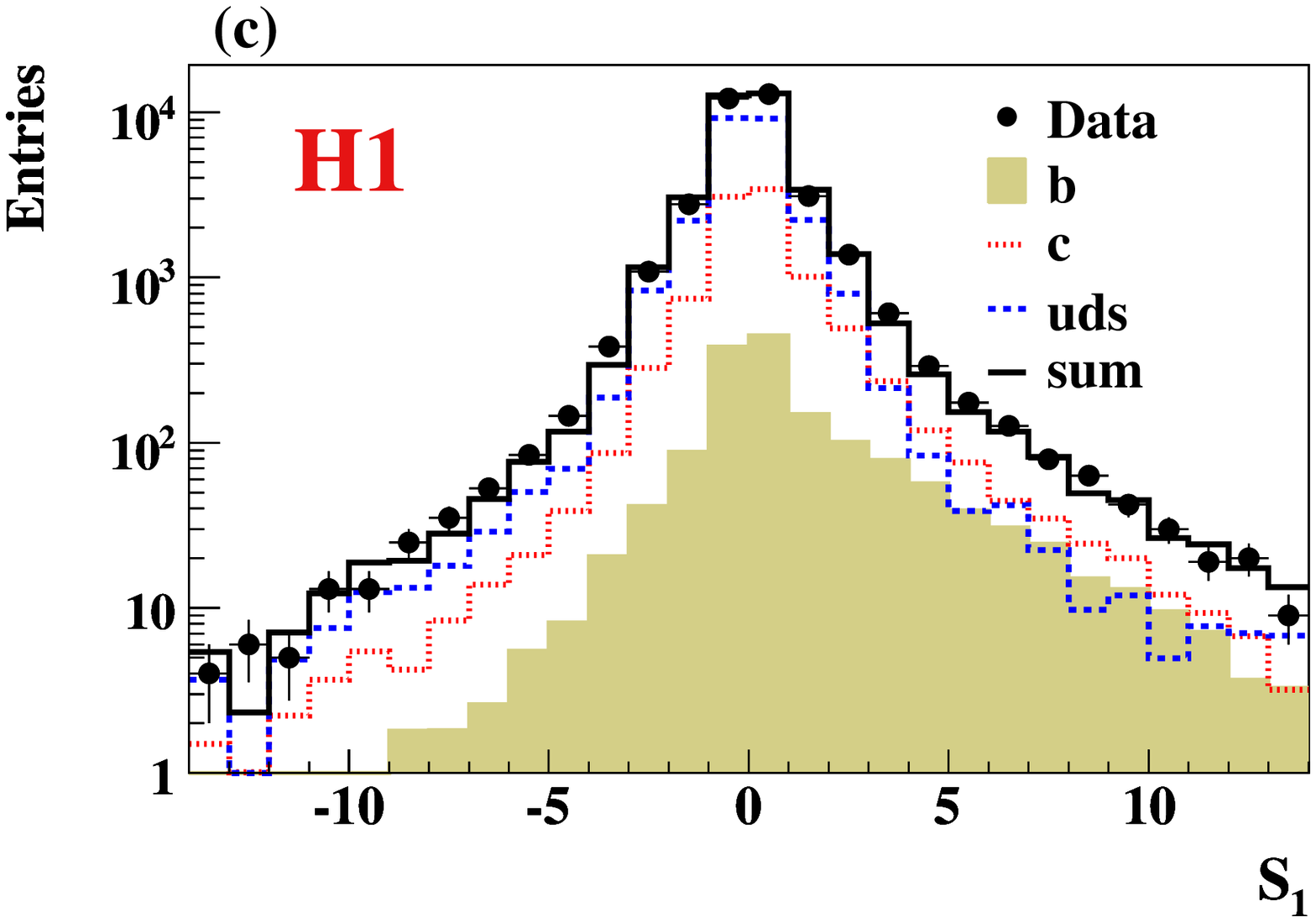}}
\put(8.1,6.){\includegraphics*[width=7.9cm]{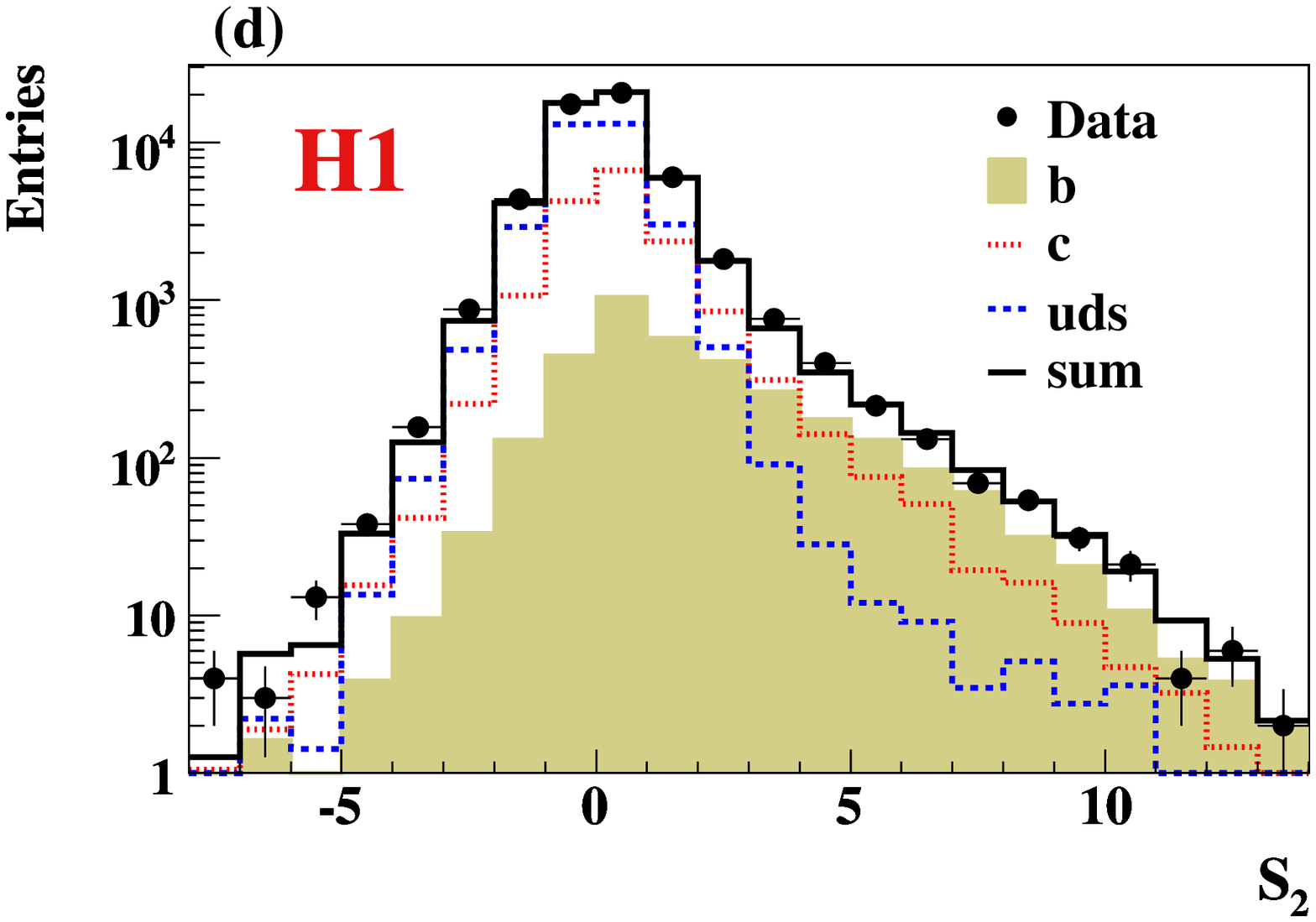}}
\put(0.,0. ){\includegraphics*[width=7.9cm]{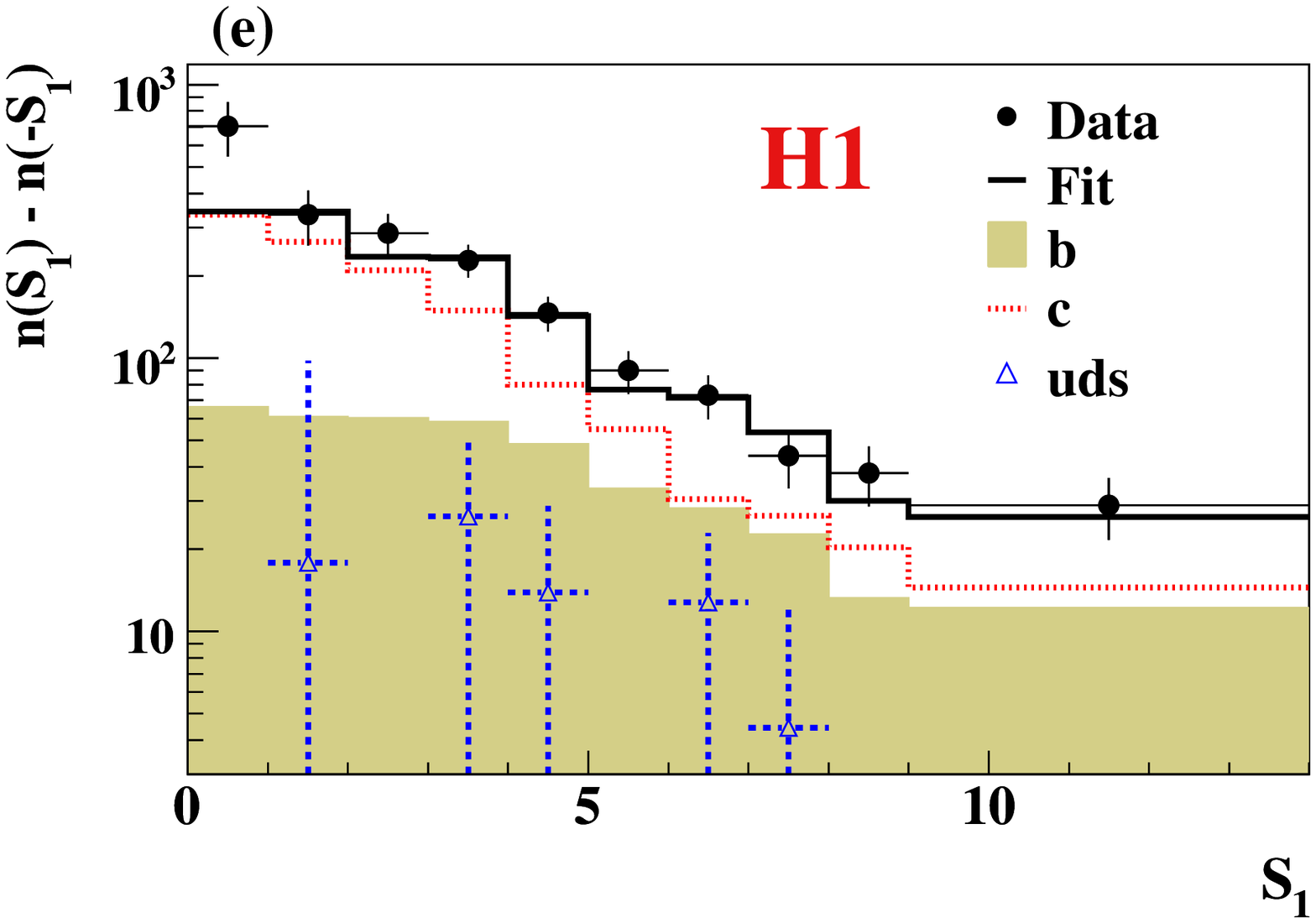}}
\put(8.1,0. ){\includegraphics*[width=7.9cm]{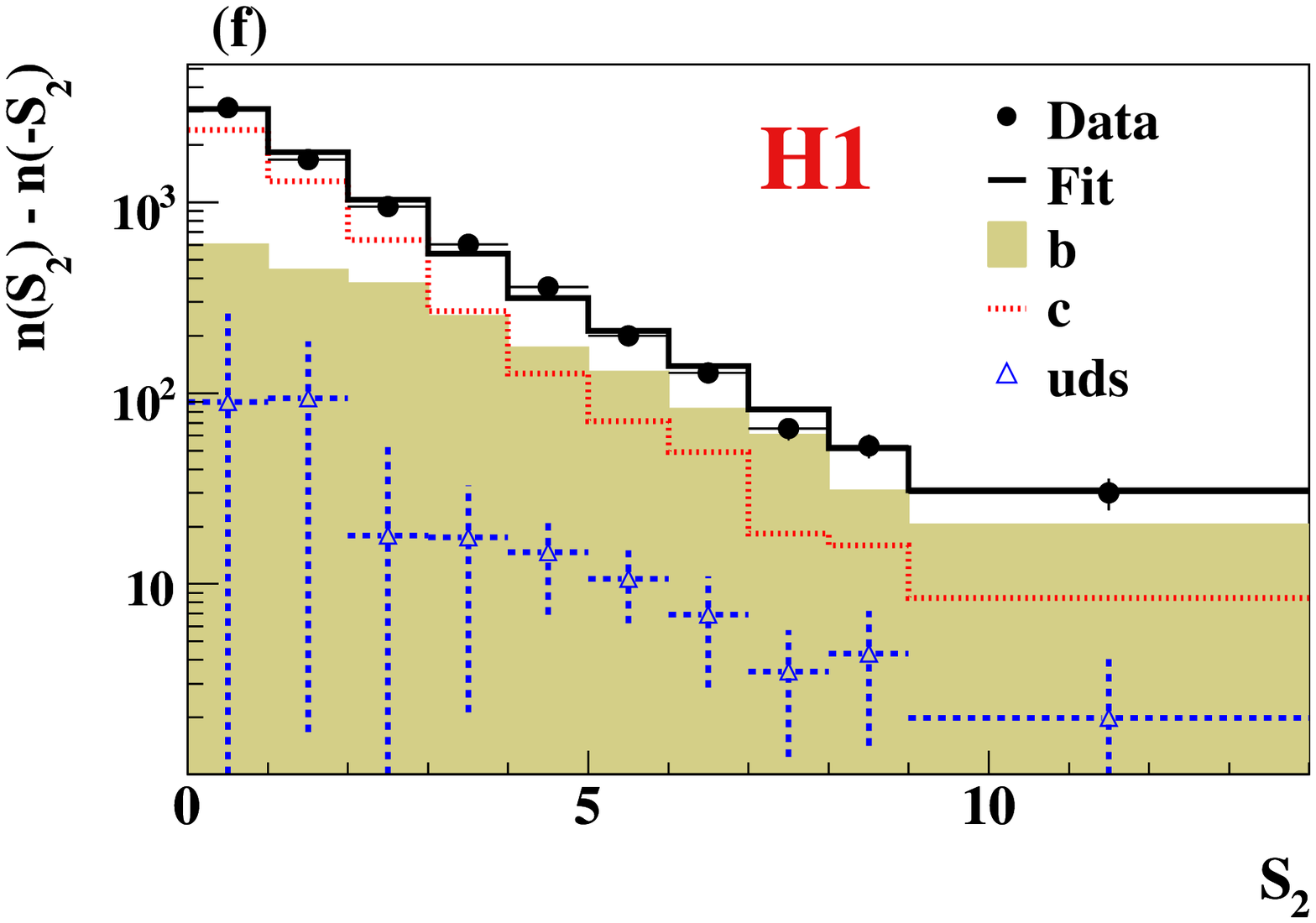}}
\end{picture}
\caption{Distributions of 
  a) the signed impact parameter $\delta$ of selected tracks, 
  b) the signed significance, 
  c) the significance $S_1$ of tracks in jets 
     with exactly one selected track, 
  d) the significance $S_2$ of the track with the second
     highest significance in jets with two or more selected tracks,
  e) the subtracted signed significance for the sample with exactly one selected
  track, 
  f) the subtracted signed significance for the sample with two or more selected
  tracks.
  In b) to f) only tracks with an impact parameter $|\delta| < 0.1$ cm 
  are considered. 
  The expectation from the PYTHIA
  Monte Carlo simulation is included in the figure, 
  showing the contributions from the various
  quark flavours after applying the scale factors obtained from the 
  fit to the subtracted significance distributions of the data
  (see section~\ref{sec:impactparm}).}
  \label{fig:delta} 
\end{figure}

The distribution of $\delta$ is shown in figure~\ref{fig:delta}a.
The data are well described by the simulation.
Due to the long lifetimes of charm and beauty flavoured hadrons
the $\delta$ distribution is asymmetric, the number of tracks with 
positive values exceeding the number of tracks with negative values. 
While the component that arises from light quarks is almost symmetric, 
the $c$ component has a moderate asymmetry and the $b$ component shows 
a marked asymmetry. 
The asymmetry seen at $|\delta|>0.1~{\rm cm}$ is mainly due to decays 
of long lived strange particles such as $K_S^0$. In order to reduce 
the effects of the strange component, tracks with $|\delta|>0.1~{\rm cm}$ 
are rejected. The significance, defined as the ratio of
the impact parameter $\delta$ to its error, is shown in
figure~\ref{fig:delta}b for all selected tracks with $|\delta|<0.1~{\rm cm}$.

The distributions of $S_1$ (figure~\ref{fig:delta}c) and
of $S_2$ (figure~\ref{fig:delta}d) show the significance of the 
selected track in jets with exactly one selected track associated to the 
jet ($S_1$) and the significance of the track with 
the second highest absolute significance in jets with two 
or more selected tracks ($S_2$).
For jets contributing to the distribution of $S_2$ it is required 
that the tracks with the first and second highest absolute significance 
in the jet have the same sign of $\delta$. At moderate and large values of 
$S_2$ the beauty contribution exceeds that from charm. 

In order to substantially reduce the uncertainty due to 
the resolution of $\delta$ and the light quark normalisation 
the negative bins in the $S_1$ and $S_2$ distributions are 
subtracted from the positive ones. 
These subtracted $S_1$ and $S_2$ distributions are shown in 
figures~\ref{fig:delta}e and f.
The distributions are dominated by charm quark events, with an increasing 
fraction of beauty quark events towards larger values of significance. 
The contribution from light quarks is seen to be small.

The $c$, $b$ and light quark fractions in the data are extracted using
a simultaneous least squares fit of simulated reference distributions
for $c$, $b$ and light quark events, obtained from the PYTHIA Monte Carlo 
simulation, to the measured subtracted $S_1$ and $S_2$ distributions 
(figures~\ref{fig:delta}e and f). The total number of events before 
any CST track selection is also used in the fit.
The Monte Carlo contributions from charm, beauty and light quark events
are scaled by factors $P_c$, $P_b$ and $P_l$ respectively, which
are the free parameters of the fit. Only the statistical errors of 
the data and the Monte Carlo simulation are taken into account. 
The fit to the complete data sample gives scale factors 
$P_c=1.45 \pm 0.14$, $P_b=1.98 \pm 0.22$, 
and $P_l=1.44 \pm 0.05$ and has a $\chi^2/n.d.f.$ of 13.1/18.
The Monte Carlo distributions shown in figures~\ref{fig:ctrl:jet1},
\ref{fig:delta}, \ref{fig:ctrl:annealing}
and~\ref{fig:ctrl:highpurity} are scaled by these factors. 
Consistent results are found when alternative 
methods are used to separate the quark flavours, such as the 
explicit reconstruction of secondary vertices described in 
section~\ref{sec:crosscheck}.

\section{Calculations in Perturbative QCD}
\label{sec:nlo}

The Monte Carlo simulation programs PYTHIA~\cite{PYTHIA} and 
CASCADE~\cite{Jung:2000hk} provide cross section predictions in 
pQCD at leading order. Parton showers are implemented 
to account for higher order effects.
PYTHIA uses the DGLAP parton evolution equations~\cite{Gribov:ri} while CASCADE
contains an implementation of the CCFM evolution equations\cite{ccfm}.

PYTHIA is run in an inclusive mode (MSTP(14)=30~\cite{PYTHIA}) 
in which direct and resolved photon processes, including heavy quark 
excitation, are generated using massless matrix elements 
for all quark flavours.
The CTEQ5L~\cite{cteq5l} parton densities are used for the 
proton and those of SaS1D~\cite{Schuler:1996fc} for the photon. 
The charm and beauty quark masses are set to 1.5 and 4.75 GeV
respectively, and the fragmentation is modelled by the Lund string 
model~\cite{Andersson:1983ia}, using the Peterson function~\cite{peterson} 
for the longitudinal fragmentation of beauty and charm quarks.

Additional Monte Carlo samples of charm, beauty and light quark events
are generated using the Monte Carlo generator CASCADE
with the charm and beauty masses as used in PYTHIA.
The process $\gamma g \rightarrow c\bar{c}$ or $b\bar{b}$ 
is implemented using off-shell matrix elements convoluted 
with $k_t$ unintegrated parton distributions in the proton.
In this analysis the parametrisation A0~\cite{Jung:2004gs} is used 
for the parton distributions.

QCD calculations to next-to-leading order are performed using the 
program FMNR~\cite{Frixione:1994dv}. 
FMNR implements the calculation at fixed order in the massive scheme, 
i.e.\,charm and beauty quarks are generated 
dynamically in the hard process via boson--gluon fusion diagrams and 
the parton distributions for the proton and the photon consist only of 
light quarks ($uds$) and gluons.
FMNR provides weighted parton level events with two or three outgoing 
partons, i.e.\,the heavy quark antiquark pair and 
possibly a third parton. 
Values of 1.5 and 4.75 GeV are chosen for the $c$ and $b$ quark masses respectively.
The renormalisation and factorisation scales are set to the transverse masses
$m_t=\sqrt{m_q^2 + p_{t,q\bar{q}}^2}$,
where $p_{t,q\bar{q}}^2$ is the average of the squared transverse momenta of 
the heavy quark and anti-quark.
The CTEQ5F3 parameterisation~\cite{cteq5l} is used for the parton 
distribution functions in the proton. 
Contributions from processes with resolved photons
are calculated using the GRV-G HO distributions of partons in the 
photon~\cite{grvg}.
In the next-to-leading order prediction these contributions are 
found to be small ($\sim 3\%$).

In order to compare the parton level calculation with the data, 
corrections from parton to hadron level are applied which are 
determined using the PYTHIA Monte Carlo event generator.
The jets at both the parton and the hadron level are reconstructed
using the inclusive $k_t$ jet-algorithm in the \pt \ recombination scheme.
The bin-by-bin corrections from parton to hadron level 
are found to be less than $\pm 10\%$ 
everywhere except in the bins $0.7<$\xgobs$<0.85$ and $0.85<$\xgobs$<1$ 
where the corrections are about $+35\%$ and $-15\%$ respectively.

Theoretical uncertainties of the NLO calculation
are estimated by independent variations of the renormalisation and 
factorisation scales by factors of one half and 
two, and the maximal changes to the cross section predictions of $30-35\%$ 
for charm and $20-30\%$ for beauty are taken as systematic errors.
The $c$ ($b$) masses are varied between 1.4 and 1.6 (4.5 and 5) GeV
leading to cross section changes of up to $\pm 4\%$. 
The cross section variations when using other proton structure 
functions such as CTEQ6M~\cite{Kretzer:2003it},
MRSG or MRST1~\cite{mrs} are less than $8\%$ in all regions of 
the measurement. The latter uncertainty is added in quadrature 
to the uncertainties from the scales and the quark mass.

\section{Cross Section Measurement}
\label{sec:crosssections}

For the measurement of the charm and beauty cross sections,
the scale factors $P_{c}$ and $P_{b}$, which are determined 
from fits of the subtracted significance distributions
to the data, are multiplied with the cross section predictions 
of the PYTHIA Monte Carlo simulation.
For the measurement of the differential charm and beauty cross sections 
the fit is performed separately in each bin $i$.
The resulting scale factors $P_{c,i}$ and $P_{b,i}$ are then multiplied 
with the bin-averaged cross section predictions of the PYTHIA Monte 
Carlo simulation, divided by the respective bin size. 
In addition, the measured differential cross sections for charm and beauty
dijet production are divided by the corresponding flavour inclusive cross 
sections to obtain the fractional contributions of events containing 
charm and beauty quarks.
The flavour inclusive dijet cross section is measured by correcting
the observed number of events before track selections
for detector efficiencies and acceptances and 
dividing by the integrated luminosity.

\subsection{Systematic Uncertainties}
\label{sec:systematics}

Systematic uncertainties of the cross section measurement are evaluated 
by variations applied to the Monte Carlo simulations.
The resulting systematic uncertainties of the total charm and 
beauty dijet cross sections are listed in table~\ref{tab:sys} and detailed below. 

\begin{table}[t]
\begin{center}
\begin{tabular}{|l|l||c|c|}
\hline 
       &           & \multicolumn{2}{c|}{Uncertainty $[\%]$} \\
Source & Variation & Charm & Beauty  \\
\hline \hline 
Impact parameter resolution  & $\oplus 25 \mu$m \; $\oplus 200\mu$m &  7  & 10  \\
Jet axis $\phi$ direction    & $1^{\circ}$ shift in $\phi$          &  3  &  2  \\
Track finding efficiency     & $2\% \oplus 1 \%$                    &  3  &  8  \\  
$uds$ asymmetry              & $\pm 50\%
$                            &      1  &  6  \\ 
\hline
HQ production model (PYTHIA) & resolved $\gamma$, \pt \ dependence  &    7  & 14  \\
Fragmentation model          & Peterson / Lund                      &    1  & 2   \\
Fragmentation fractions      & PDG                                  &  0.5  & 1.6 \\
Hadron lifetimes             & PDG                                  &  0.1  & 0.3 \\
Charged track multiplicities & MARK-III, LEP, SLD                   &  1.5  & 4   \\
\hline
Jet energy scale             & $2\%$                                &    6  & 5   \\ 
Trigger efficiency           &                                      &    5  & 5   \\
Luminosity measurement       &                                      &  1.5  & 1.5 \\
\hline                   
\hline                   
Total                        &                                      &  14   & 22  \\ 
\hline
\end{tabular}
\caption{Systematic uncertainties of the measured total charm and beauty dijet 
cross sections.}   
\label{tab:sys}
\end{center}
\end{table}

\begin{itemize}
\item An uncertainty in the $\delta$ resolution of the tracks is estimated 
  by varying the resolution by an amount that encompasses the differences 
  between the data and simulation (figure~\ref{fig:delta}).
  This is achieved by applying an additional Gaussian smearing in the Monte
  Carlo simulation of 200~$\mu{\rm m}$ to $5\%$ of randomly selected tracks 
  and of 25~$\mu{\rm m}$ to the rest. 
\item The uncertainty of the jet axis reconstruction is estimated by 
  shifting the jet axis in 
  azimuth $\phi$ by $\pm 1^\circ$.
\item The reconstruction efficiency of central drift chamber tracks 
  is uncertain to the level of $2\%$ and the efficiency for these
  tracks to have hits in both $r$-$\phi$ layers of the silicon vertex detector 
  is known to $1\%$.   
\item The uncertainty resulting from the shape of the subtracted
  significance
  distributions $S_1$ and $S_2$ for light quarks 
  (figures~\ref{fig:delta}e and f) is estimated by repeating the fits 
  with the light quark $S_1$ and $S_2$ distributions varied by
  $\pm 50\%$ of the default value.   
\item The systematic error arising from the uncertainty of the underlying
  physics model is estimated by varying the contribution from resolved photon
  processes in the PYTHIA prediction by $\pm 50\%$, and by reweighting 
  the \pt \ distribution as predicted by PYTHIA 
  to that of CASCADE. These variations lead to changes of the cross sections of
  $\pm 7\%$ for charm and $\pm 14\%$ for beauty.
\item The uncertainties in the description of the heavy quark fragmentation 
  are estimated by repeating the fits with Monte Carlo simulation templates
  in which the Lund Bowler function~\cite{Andersson:1983ia} 
  is used for the longitudinal fragmentation instead of the Peterson 
  function. 
\item The uncertainties arising from the various $D$ and $B$ hadron lifetimes,
  fragmentation fractions and mean charged track multiplicities are
  estimated by varying the input values of the Monte Carlo simulation by the 
  experimental errors of the corresponding measurements or world averages.
  For the fragmentation fractions of $c$ and $b$ quarks 
  to hadrons and for the lifetimes of these hadrons the central values and errors 
  on the world averages are taken from~\cite{Eidelman:2004wy}. For the mean 
  charged track multiplicities the values and uncertainties for $c$  
  hadrons are taken from Mark-III~\cite{Coffman:1991ud} and for $b$ hadrons
  from LEP and SLD measurements~\cite{lepjetmulti}.
\item The uncertainty of the jet energy scale of $2\%$ leads to 
  cross section uncertainties between $3\%$ at small $p_{\textrm{t}}^{\textrm{jet}}$ 
  and $12\%$ at large $p_{\textrm{t}}^{\textrm{jet}}$ and $6\%$ on average, 
  independently of the quark flavour.
\item The trigger efficiency is studied using monitoring events from neutral 
  current processes in deep inelastic scattering in which the scattered positron 
  triggers the events independently of the triggers under study. The uncertainty is 
  determined to be $5\%$.
\item The luminosity is known to an accuracy of $1.5\%$.  
\end{itemize}

Total systematic uncertainties of $14\%$ and $22\%$ 
are obtained for the measurement of the charm and 
the beauty production cross sections respectively.
The total systematic error for the flavour inclusive dijet cross section
is $8\%$ resulting from the uncertainty of the hadronic 
energy scale ($6\%$), the trigger efficiency uncertainty ($5\%$) 
and the uncertainty of the luminosity measurement ($1.5\%$).
For the relative contributions of charm and beauty production to the flavour
inclusive dijet cross section, the statistical errors are added in
quadrature and the systematic errors include those sources
that are specific to the charm and beauty cross section measurement.
The same uncertainties are equally attributed to all bins of the 
measurement except for the uncertainty of the hadronic energy scale
for which the uncertainties are determined and applied individually 
in each bin of the measurement. 

\subsection{Results}
\label{sec:results}

The total dijet charm photoproduction cross section in the range
$Q^2<1$ GeV$^2$, $0.15<y<0.8$, 
$\ptjetsm>11(8)$ GeV and \mbox{ $-0.9<\eta^{\textrm{jet}_{1(2)}}<1.3$}
is measured to be 
$$
\sigma(ep\rightarrow ec\bar{c}X \rightarrow ejjX) = 
702 \pm 67 (stat.) \pm 95 (syst.) {\rm pb}.    
$$
For the same kinematic range, the measured beauty cross section is
$$
\sigma(ep\rightarrow eb\bar{b}X \rightarrow ejjX) = 
150 \pm 17 (stat.) \pm 33 (syst.) {\rm pb}.    
$$
The predictions from the theoretical calculations 
are detailed in table~\ref{tab:sigma}.
The NLO parton level calculations are corrected to the hadron 
level using correction factors as determined from PYTHIA (see section~\ref{sec:nlo}).
For charm, the NLO QCD calculation FMNR is somewhat lower than the measurement
but still in reasonable agreement within the theoretical errors.
For beauty, FMNR is lower than the data by a factor 1.8, corresponding
to 1.6 standard deviations, taking both experimental uncertainties and 
uncertainties in the theory into account. 
For both, charm and beauty, PYTHIA and CASCADE predict a normalisation 
which is similar to that of FMNR.

\begin{table}
\begin{center}
\begin{tabular}{|r||c|c|}
\hline 
            & Charm [pb]                           & Beauty [pb] \\ \hline
Data        & $702 \pm 67 (stat.) \pm 95 (syst.) $ & $150 \pm 17 (stat.) \pm 33 (syst.) $ \\
FMNR        & $500 ^{+173}_{-99}$                  & $83 ^{+19}_{-14}$ \\
PYTHIA      & $484$                                & $76$ \\
CASCADE     & $438$                                & $80$ \\
\hline
\end{tabular}
\caption{The measured charm and beauty photoproduction dijet cross sections
in the kinematic range $Q^2<1$ GeV$^2$, $0.15<y<0.8$, 
$\ptjetsm>11(8)$ GeV and $-0.9<\eta^{jet_{1(2)}}<1.3$ in comparison
to predictions in NLO QCD (FMNR) and from the Monte Carlo programs PYTHIA and CASCADE.}
\label{tab:sigma}
\end{center}
\end{table}

\begin{figure}[htb]
\unitlength1cm
\begin{picture}(8,18)
\put(-.25,12.){\includegraphics*[width=8.1cm]{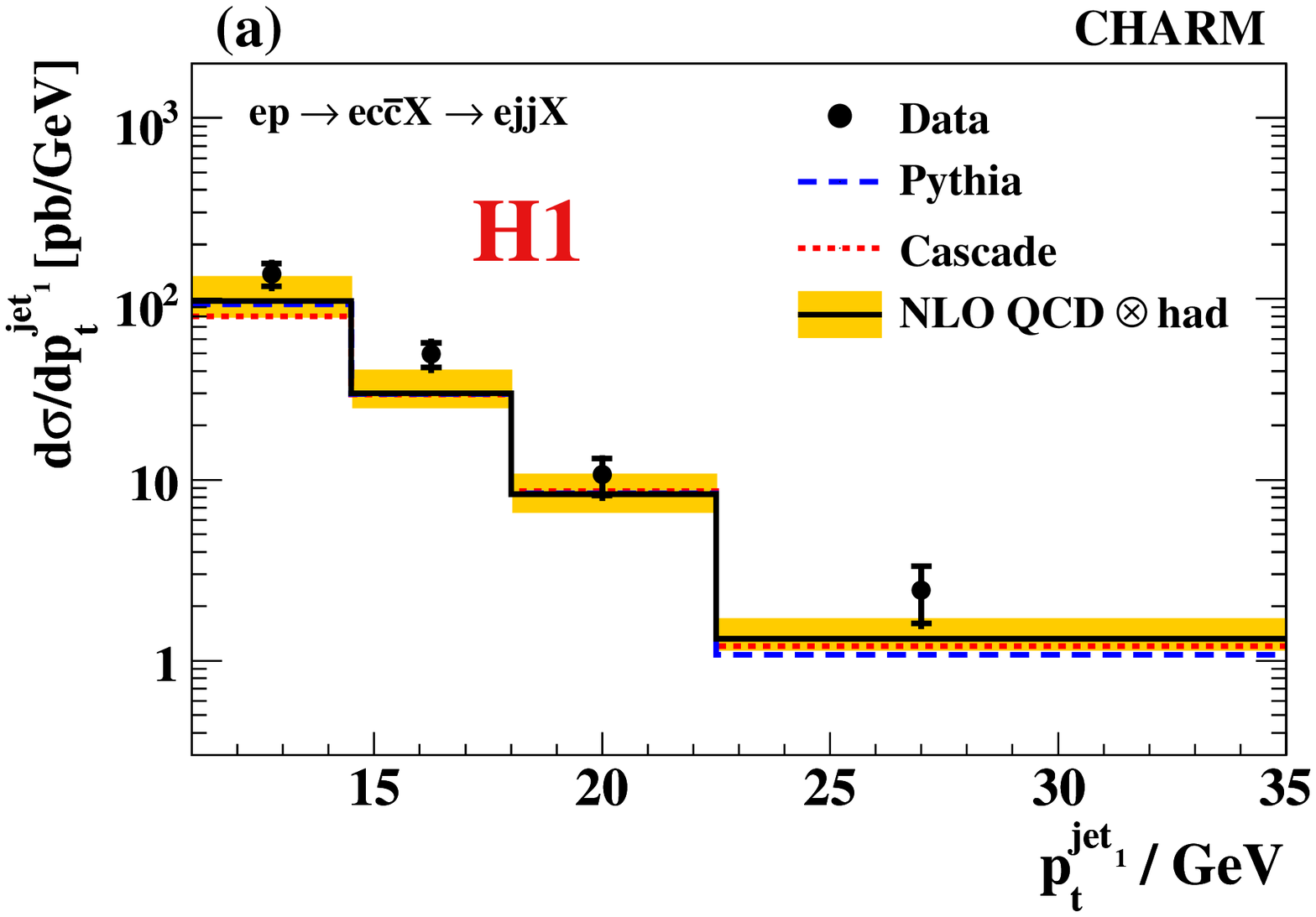}}
\put(8.05,12.){\includegraphics*[width=8.1cm]{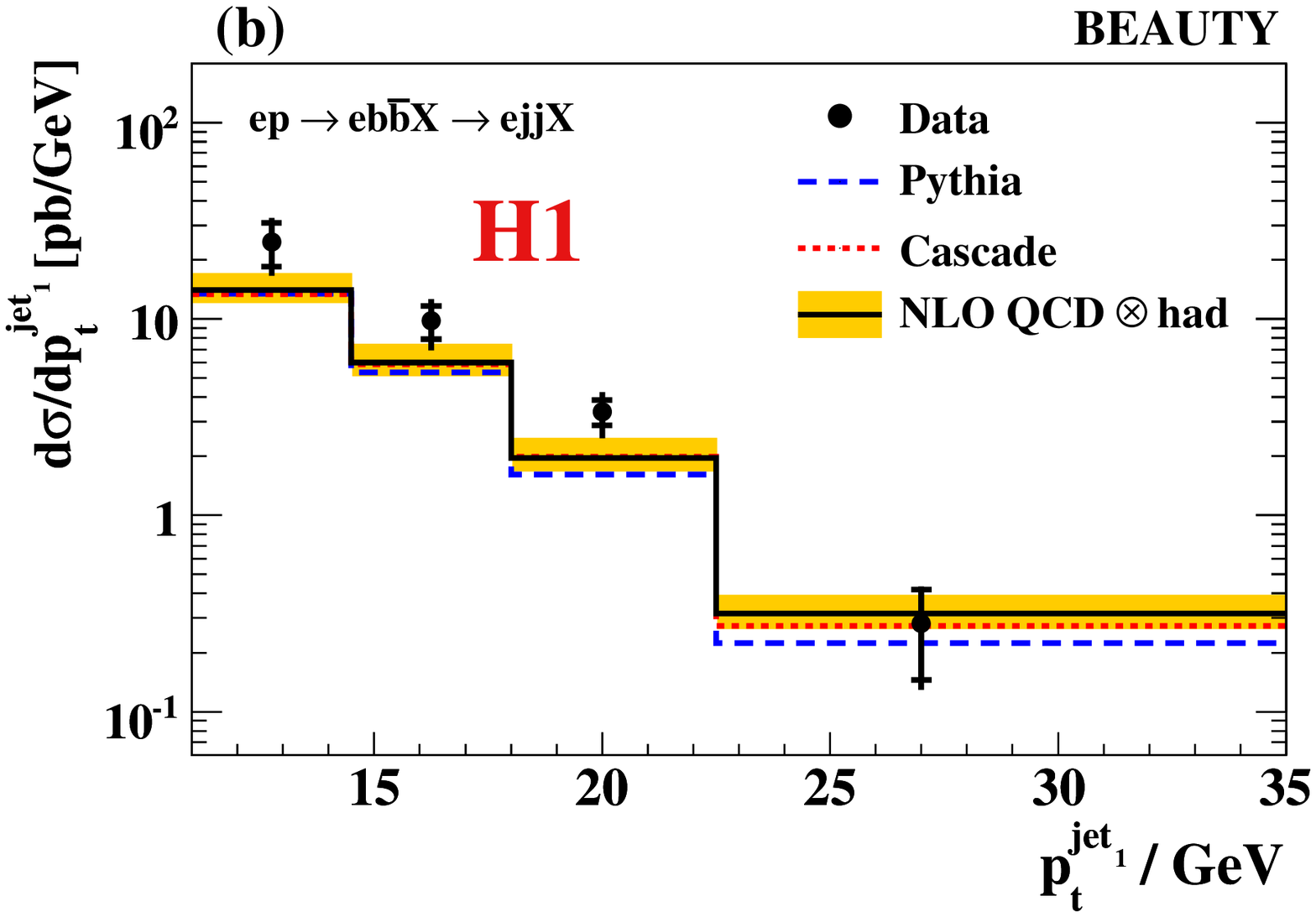}}
\put(-0.1,6.) {\includegraphics*[width=7.82cm]{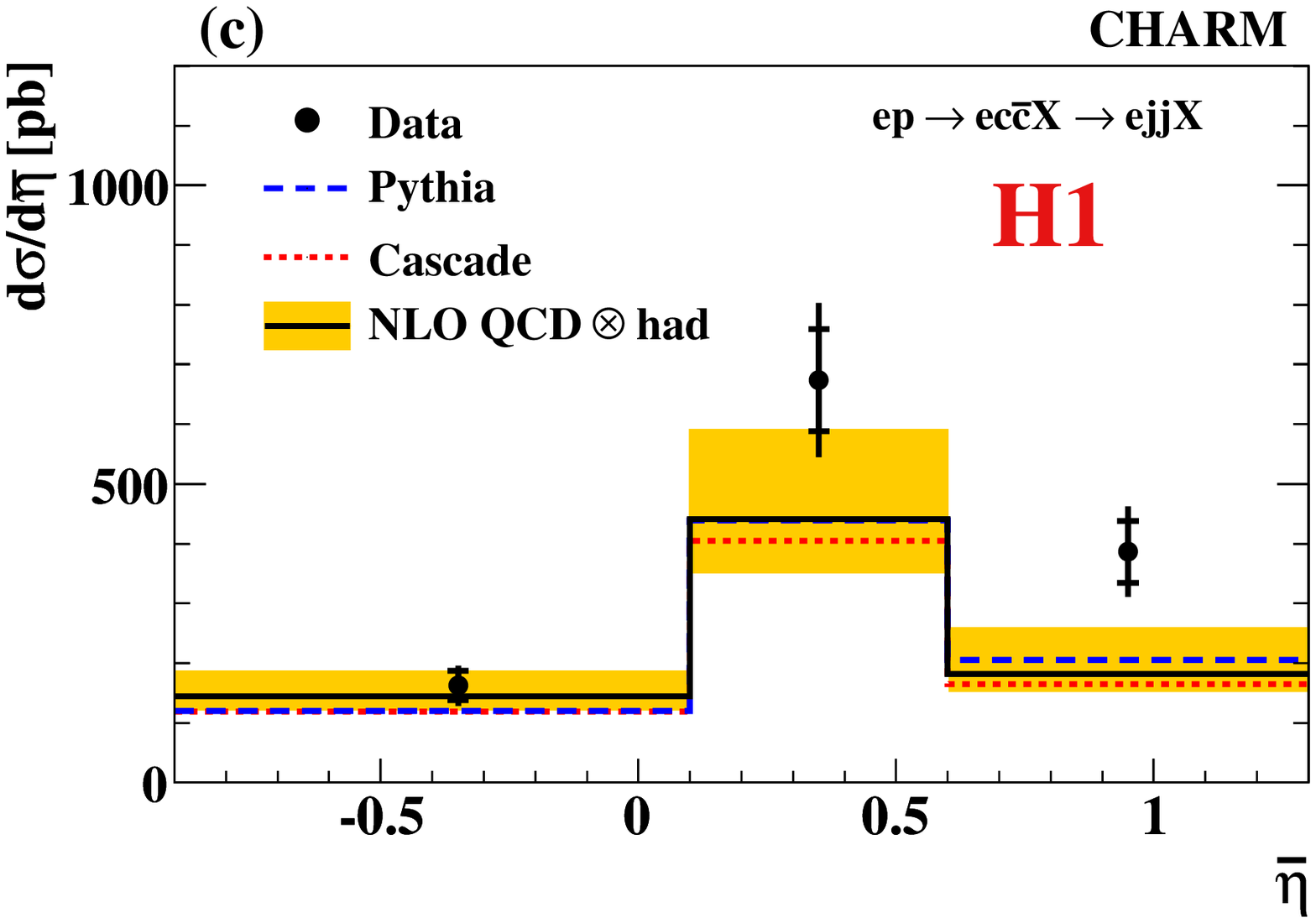}}
\put(8.2,6.){\includegraphics*[width=7.82cm]{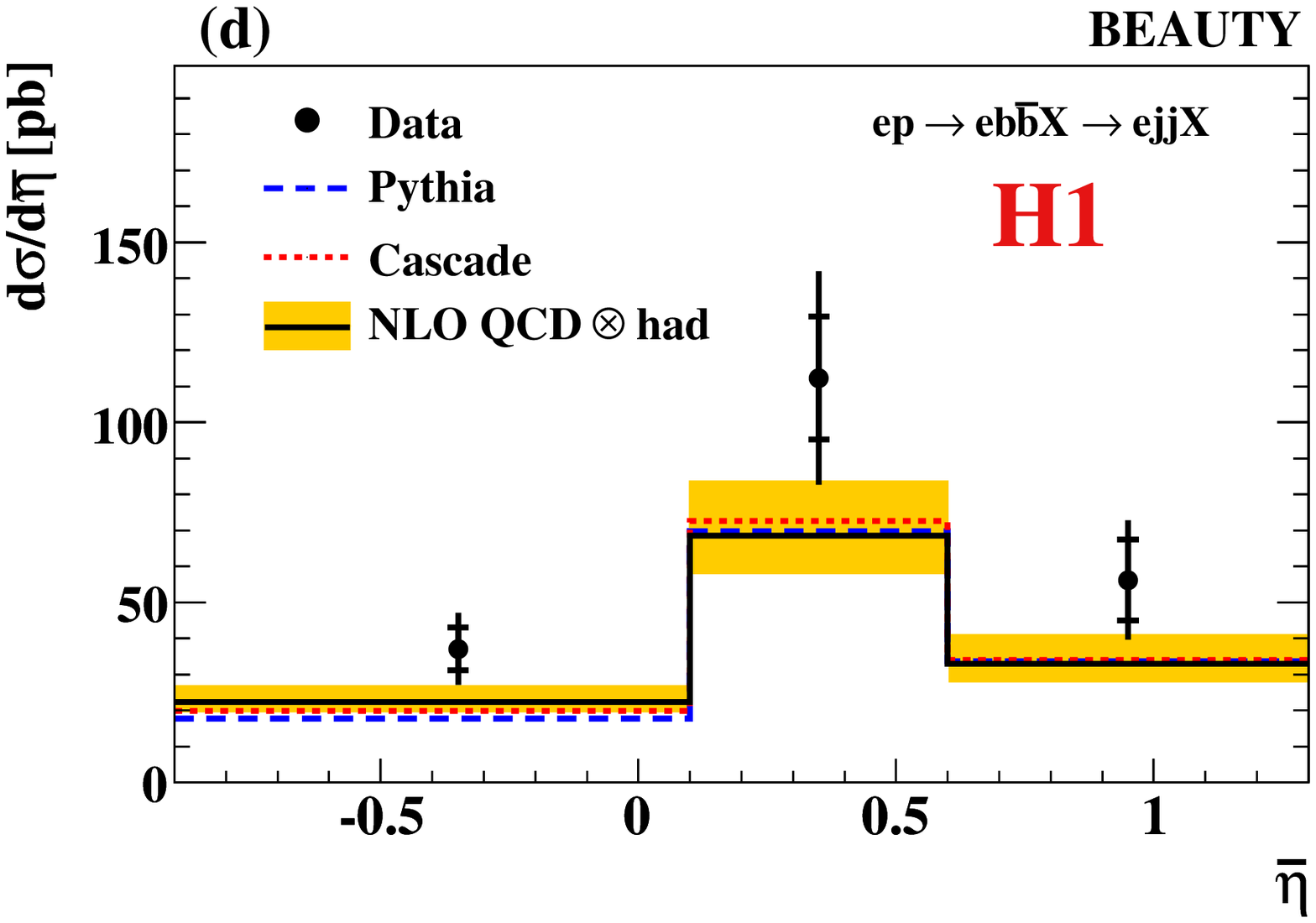}}
\put(-0.1,0. ){\includegraphics*[width=7.9cm]{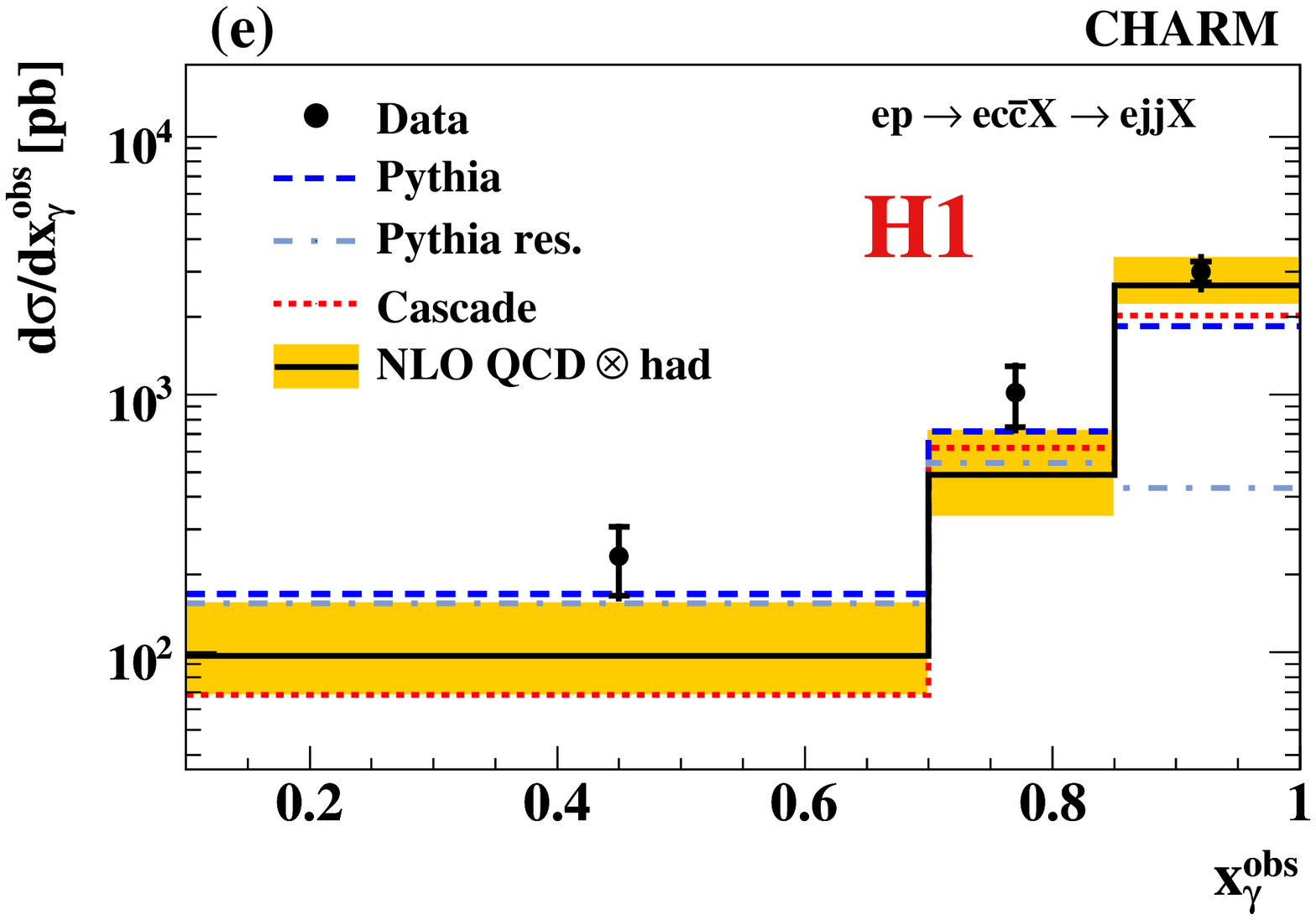}}
\put(8.15,0. ){\includegraphics*[width=7.9cm]{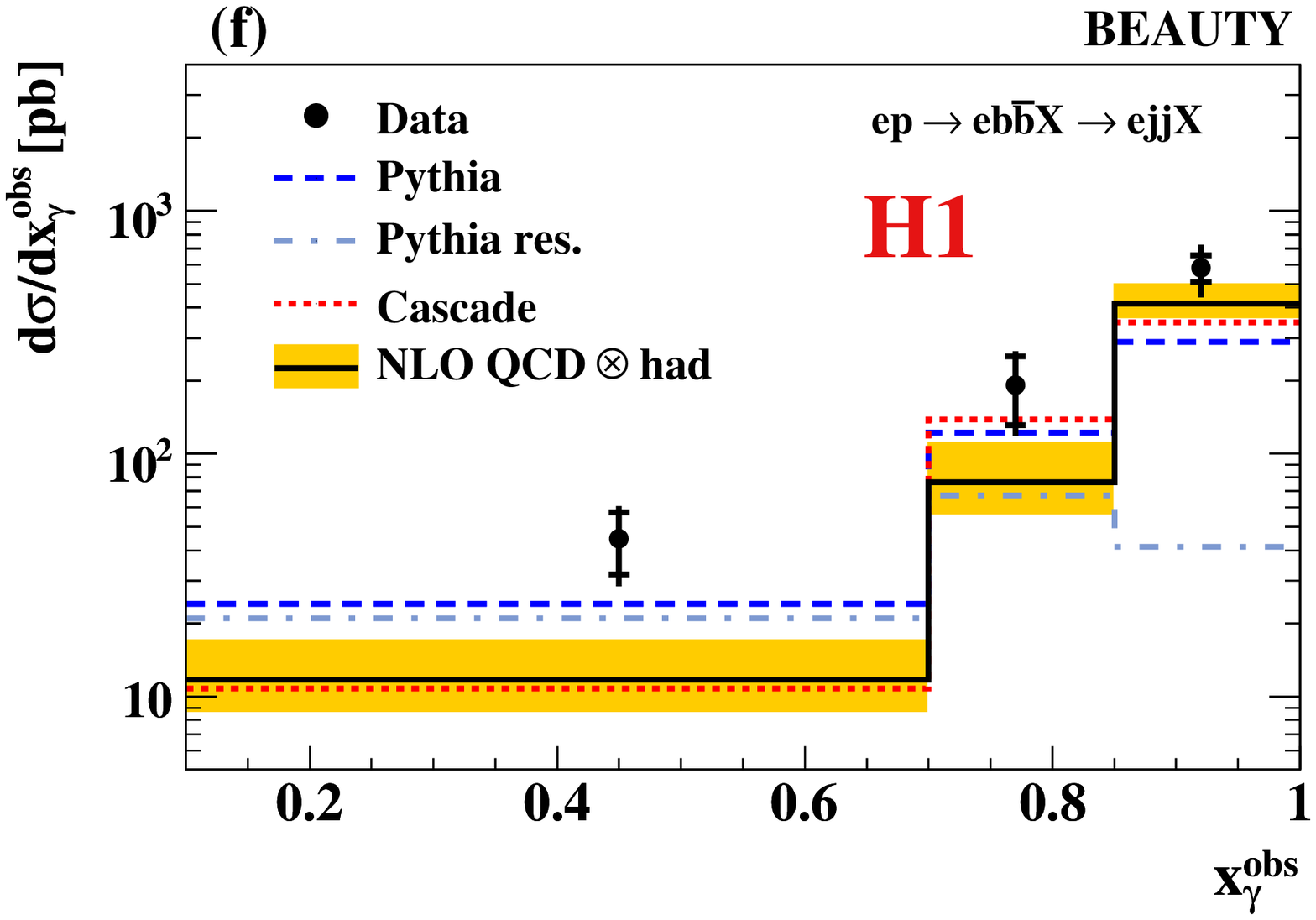}}
\end{picture}
    \caption{Differential charm and beauty photoproduction cross sections 
    a,b) $d\sigma/d\ptjetm$,
    c,d) $d\sigma/d\bar{\eta}$ and
    e,f) $d\sigma/d\xgobsm$ for the process
    $ep\rightarrow e(c\bar{c}~~{\rm or}~~b\bar{b})X \rightarrow ejj X$.
    The inner error bars indicate the statistical
    uncertainty and the outer error bars show the statistical and systematic
    errors added in quadrature.
    The solid lines indicate the prediction from a NLO QCD calculation,
    corrected for hadronisation effects,
    and the shaded band shows the estimated uncertainty.
    The absolute predictions from PYTHIA (dashed lines) and
    CASCADE (dotted lines) are also shown. The contribution from 
    resolved processes in PYTHIA (dash-dotted lines) is depicted separately.}
\label{fig:xsecs:cb} 
\end{figure}

The measured differential cross sections as functions of \ptjet, of
$\bar{\eta}$ and of \xgobs, are shown in figure~\ref{fig:xsecs:cb} and listed 
in table\,\ref{tab:xsecs:inclusive}.  PYTHIA is used to 
determine the point in the bin at which the bin-averaged cross section 
equals the differential cross section.
Both the charm and beauty data are reasonably well described in shape.
A large difference, however, between 
the beauty data and the NLO QCD calculation is observed in the region 
of small values of \xgobs \ (figure~\ref{fig:xsecs:cb}f), where the prediction 
lies significantly below the data. In this region, PYTHIA
predicts a large contribution from events with resolved photons, as 
indicated in the figures by the dashed-dotted lines. 
PYTHIA describes the shapes of the charm and beauty data 
distributions, while the normalisations are low. 
The CASCADE prediction is too small in the region 
of small $\xgobsm$, but approaches the measurement in the region
$\xgobsm>0.85$.

\begin{figure}[htb]
\unitlength1cm
\begin{picture}(8,12)
\put(-.25,6.){\includegraphics*[width=8.1cm]{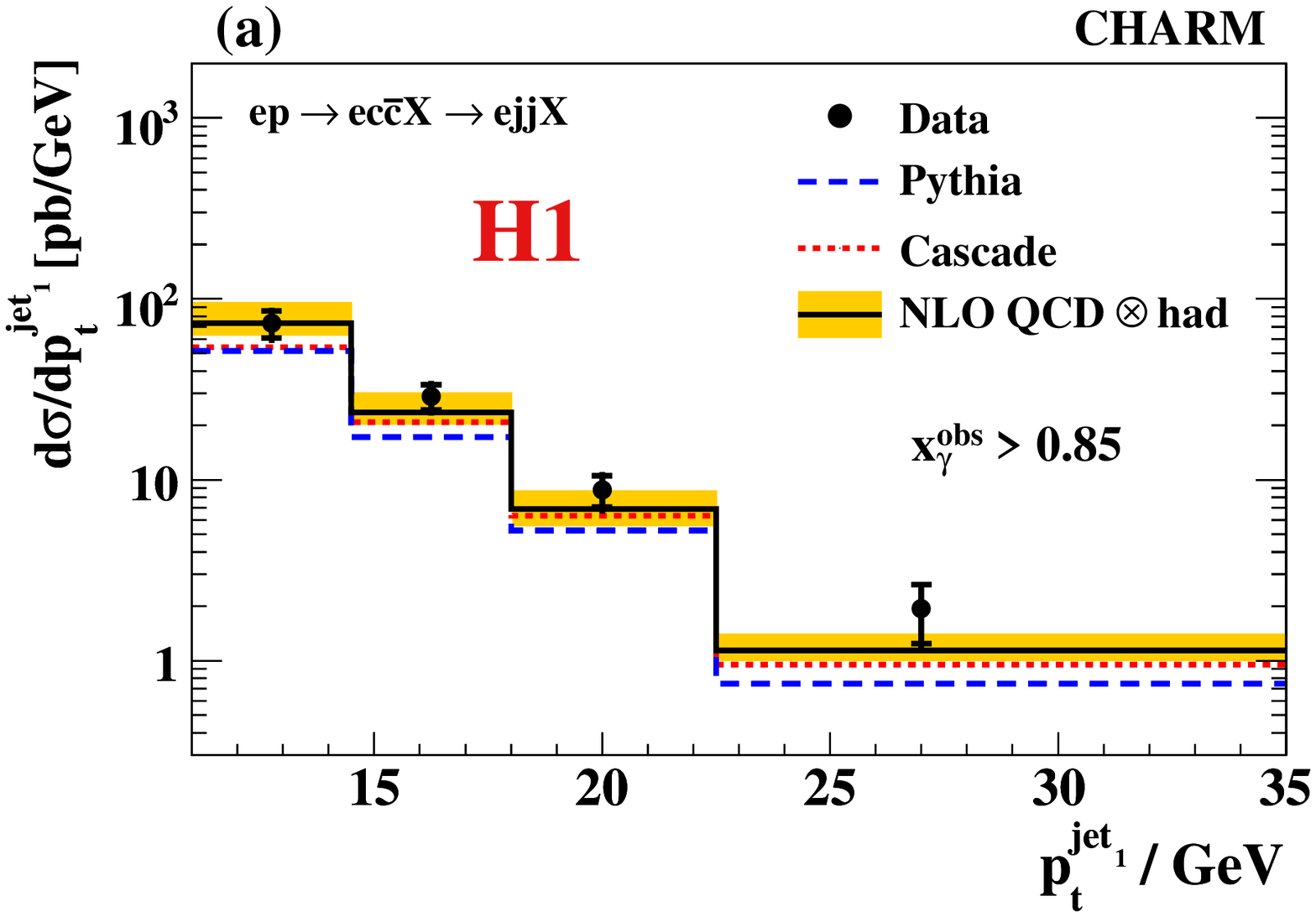}}
\put(8.05,6.){\includegraphics*[width=8.1cm]{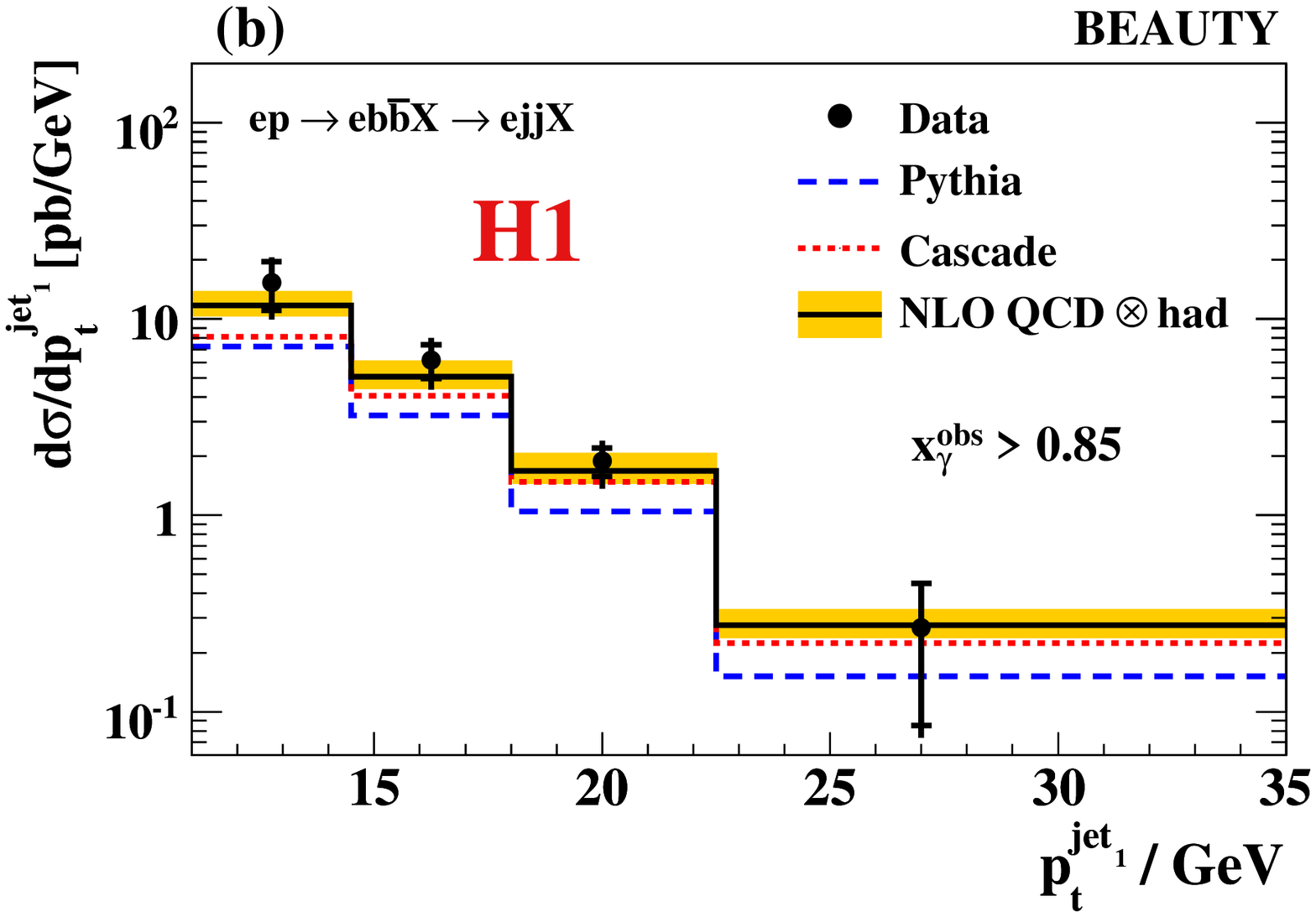}}
\put(-.1,0.) {\includegraphics*[width=7.82cm]{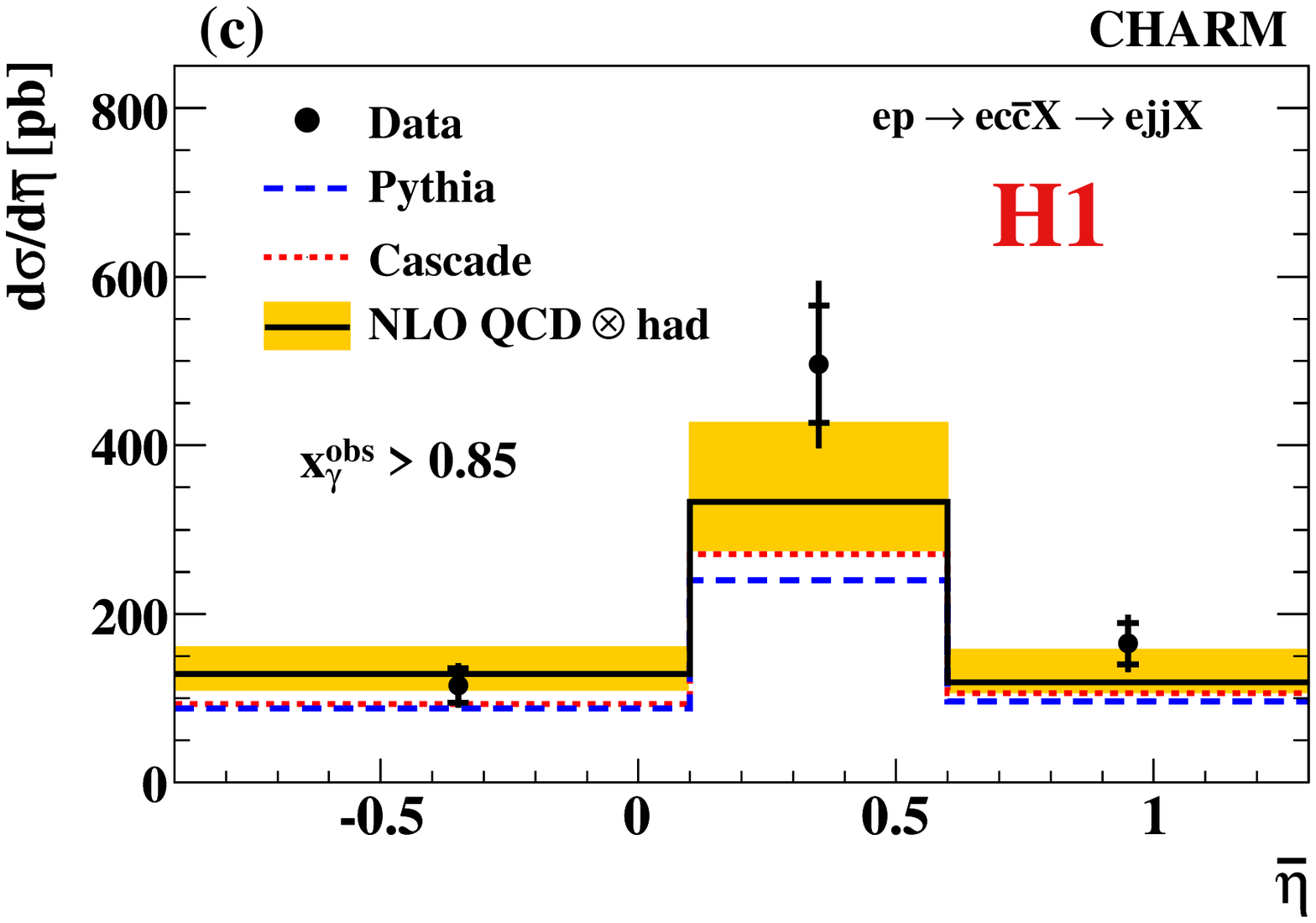}}
\put(8.2,0.){\includegraphics*[width=7.82cm]{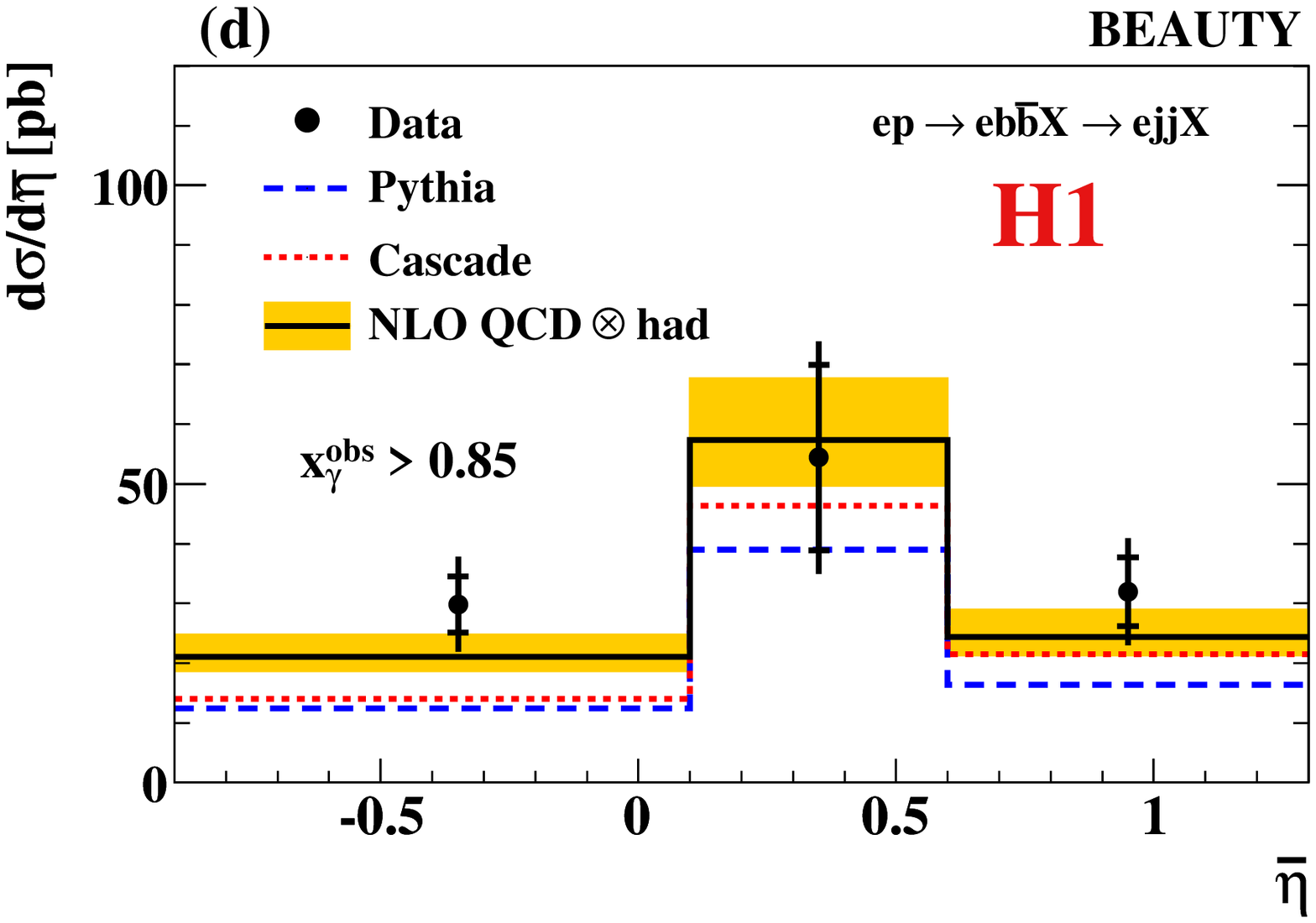}}
\end{picture}
    \caption{Differential charm and beauty photoproduction cross sections 
    a,b) $d\sigma/d\ptjetm$ and
    c,d) $d\sigma/d\bar{\eta}$ for the process
    $ep\rightarrow e(c\bar{c}~~{\rm or}~~b\bar{b})X \rightarrow ejj X$
    in the region $\xgobsm>0.85$.
    The inner error bars indicate the statistical
    uncertainty and the outer error bars show the statistical and systematic
    errors added in quadrature.
    The solid lines indicate the prediction from a NLO QCD calculation,
    corrected for hadronisation effects,
    and the shaded band shows the estimated uncertainty.
    The absolute predictions from PYTHIA (dashed lines) and
    CASCADE (dotted lines) are also shown.}
\label{fig:xsecs:cb_085} 
\end{figure}

Differential cross sections are also measured 
separately for the region $\xgobsm>0.85$ and the
results as functions of \ptjet \ and of $\bar{\eta}$ are
shown in figure~\ref{fig:xsecs:cb_085} and listed in 
table~\ref{tab:xsecs:direct}.
The charm cross sections are in good agreement 
with the NLO QCD calculation both in normalisation and shape. 
The beauty cross sections are also reasonably well described, the agreement being
significantly better than for the whole range of $\xgobsm$. 

\begin{figure}[htb]
\unitlength1cm
\begin{picture}(8,18)
\put(3.5,12. ){\includegraphics*[width=8.5cm]{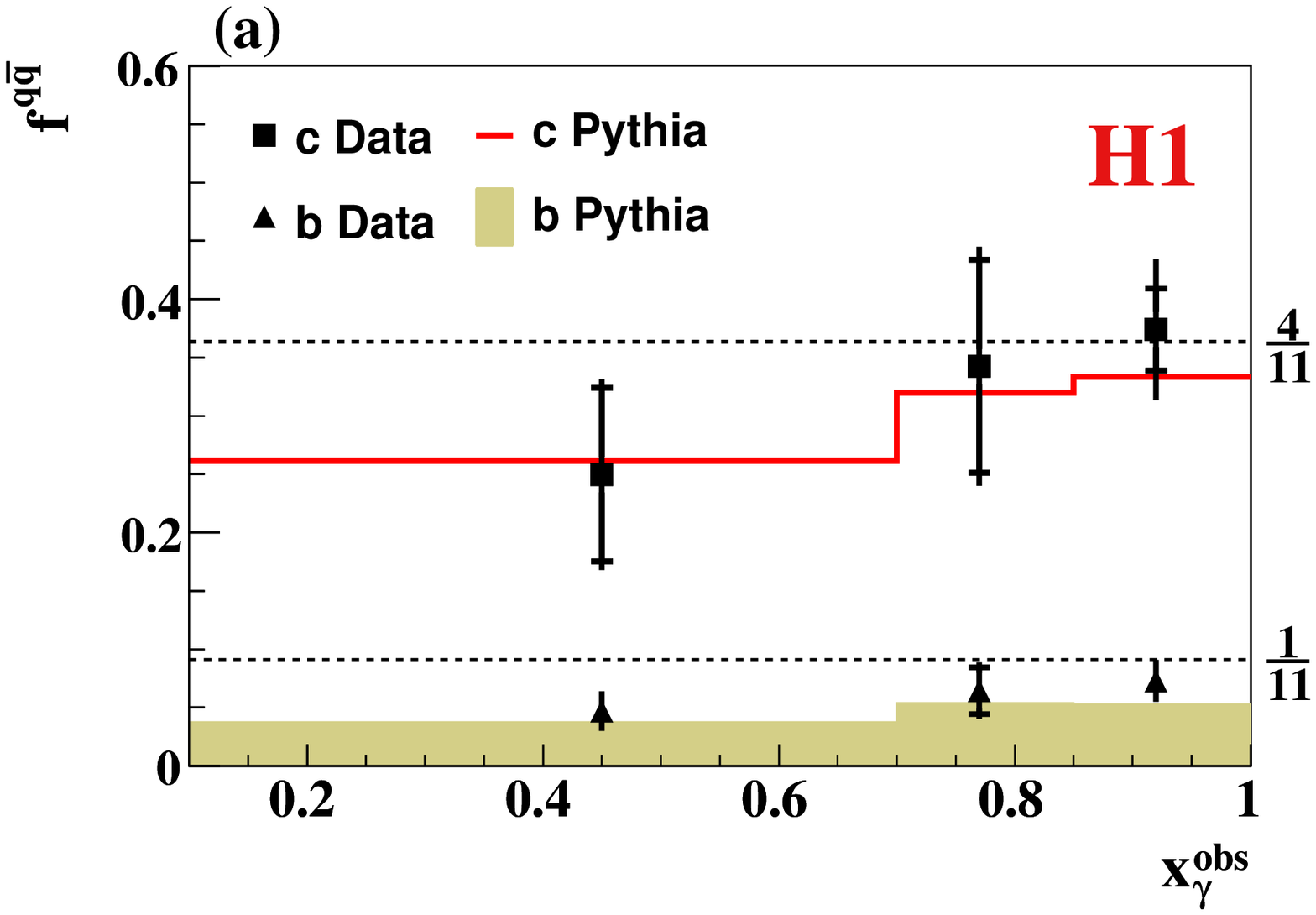}}
\put(3.5,6.){\includegraphics*[width=8.5cm]{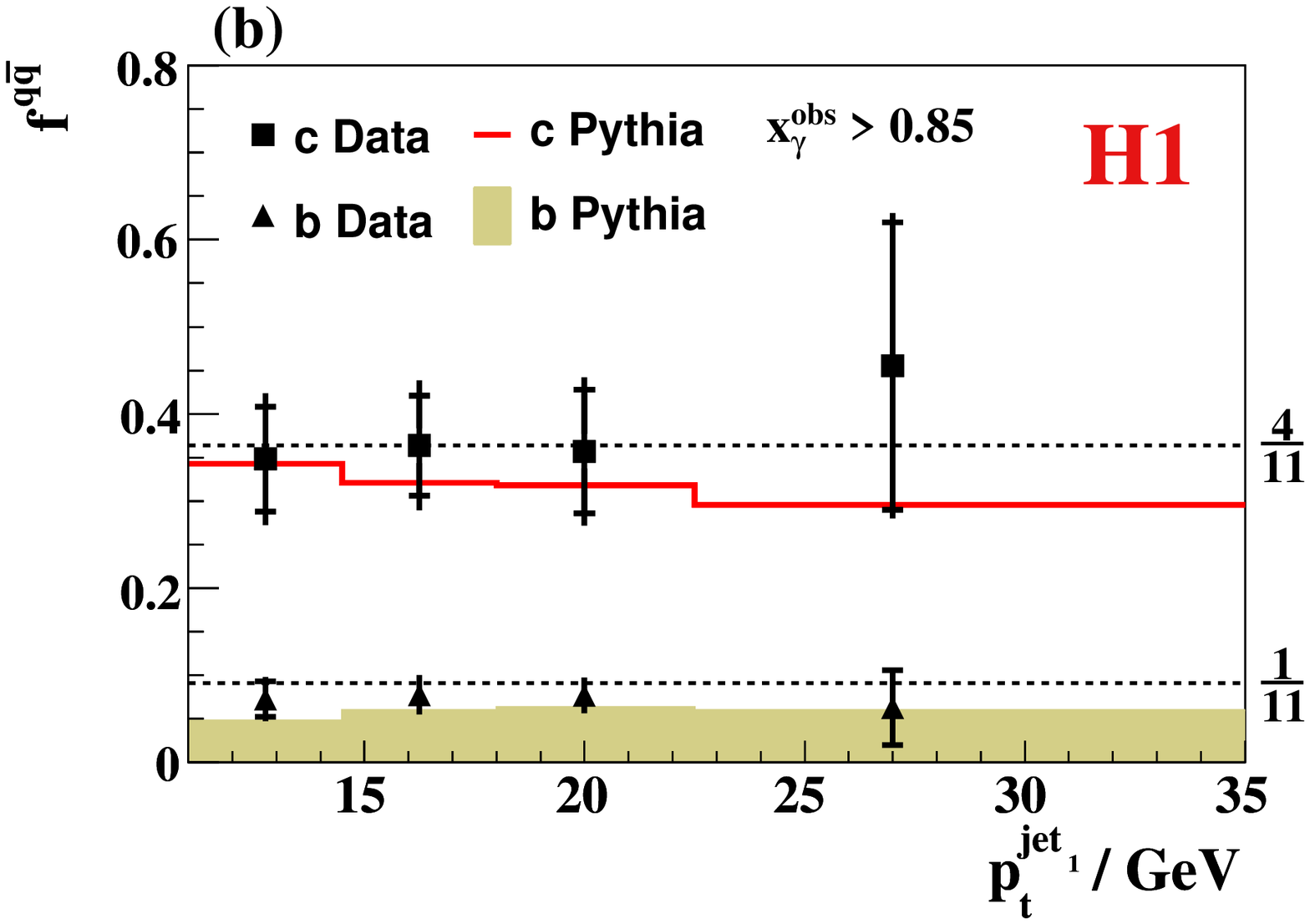}}
\put(3.5,0.){\includegraphics*[width=8.5cm]{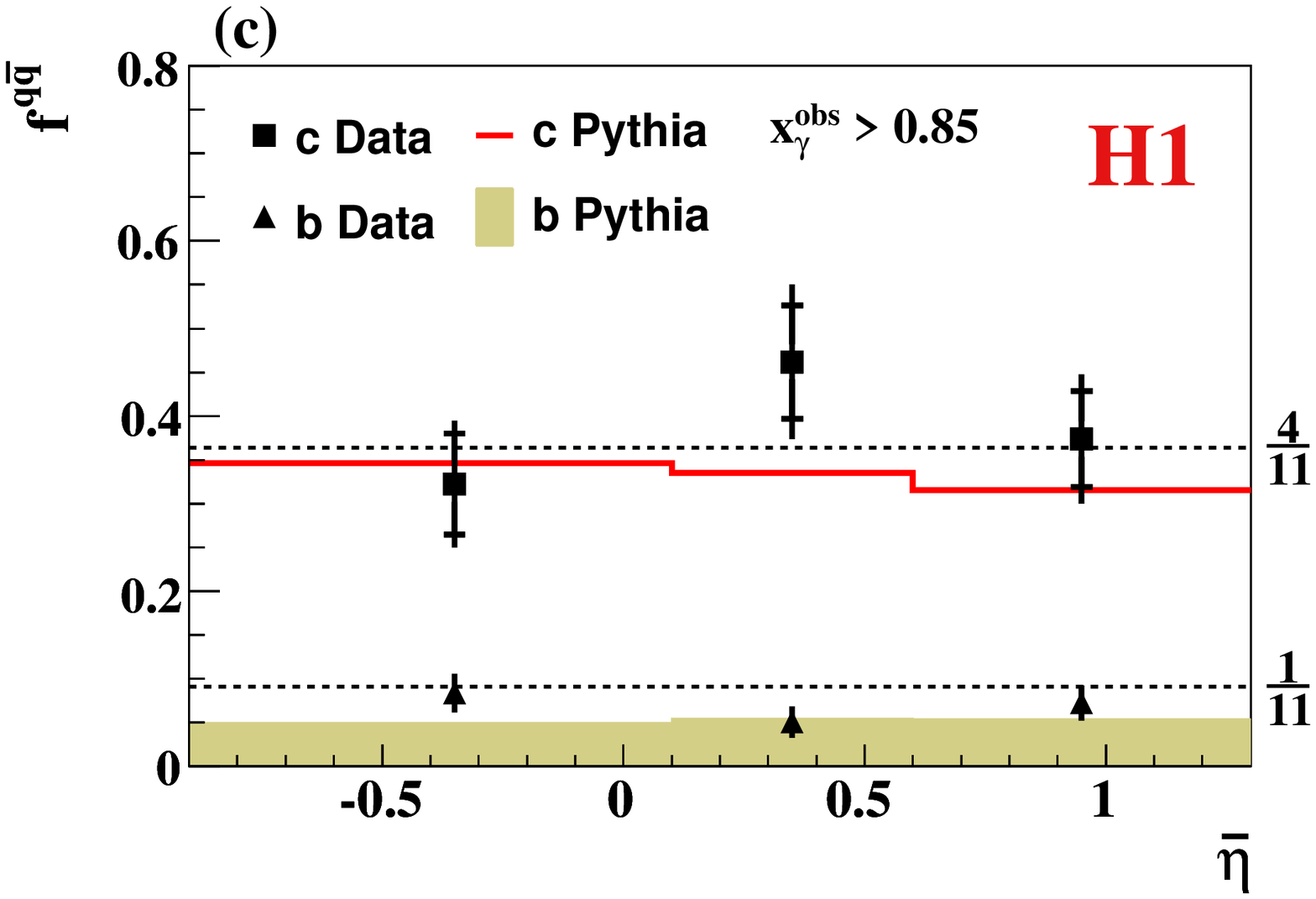}}

\end{picture}
    \caption{Relative contributions from charm (squares) and beauty events (triangles)
    as a function of 
    a) the observable $\xgobsm$,
    b) the transverse momentum \ptjet \ of the leading jet for $\xgobsm>0.85$ 
    and c) the mean pseudo-rapidity $\bar{\eta}$ of the two jets for $\xgobsm>0.85$. 
    The inner error bars indicate the statistical
    uncertainty and the outer error bars show the statistical and systematic
    error added in quadrature.     
    The solid line (shaded area) indicates the absolute 
    prediction from PYTHIA for charm (beauty).}
  \label{fig:xsecs:incl} 
\end{figure}

The relative contributions from charm and beauty 
to the inclusive dijet cross sections are presented
in figure~\ref{fig:xsecs:incl} and listed 
in tables\,\ref{tab:xsecs:inclusive} and~\ref{tab:xsecs:direct}.
In figure~\ref{fig:xsecs:incl}a, the
relative contributions are shown as a function of $\xgobsm$.
The data are compared with the PYTHIA Monte Carlo simulation which
predicts an increase of the relative charm and beauty contributions
towards large $\xgobsm$ where direct photon-gluon fusion processes dominate.

Assuming the charm and beauty quarks to be light,
na\"ive quark charge counting predicts a value of four
for the relative production rates of charm to beauty dijets 
in direct photon-gluon fusion processes.
In comparison, the measurement in the region $\xgobsm>0.85$
yields a ratio of 5.1 $\pm$ 1.1 (stat.). 
In figure~\ref{fig:xsecs:incl}b and c 
the relative contributions to the dijet cross section are shown for 
the region $\xgobsm>0.85$ as a function of \ptjet \ and 
of $\bar{\eta}$. The ratios are constant within their uncertainties.

\section{Heavy Quark Enriched Data Sample}
\label{sec:crosscheck}

In order to study the description of the data by the PYTHIA Monte Carlo
further, a subsample of events is used, in which the fraction of events 
containing heavy quarks is enriched.
For this subsample secondary vertices of the heavy hadron decays are
reconstructed explicitly using events with
at least two selected tracks in a given jet. 
The relative fraction of heavy quarks in the event sample is then 
further enhanced by subsequent cuts on the secondary vertex track 
multiplicity and the decay 
length significance. For this heavy quark enriched sample the distributions 
of the jet transverse momentum, the mean pseudo-rapidity and the observable 
$\xgobsm$ are presented and the decomposition and shape differences 
of direct and resolved photon processes, as implemented in PYTHIA, are studied.

\begin{figure}[htb]
\unitlength1cm
\begin{picture}(8,18)
\put(3.5,12. ){\includegraphics*[width=8.5cm]{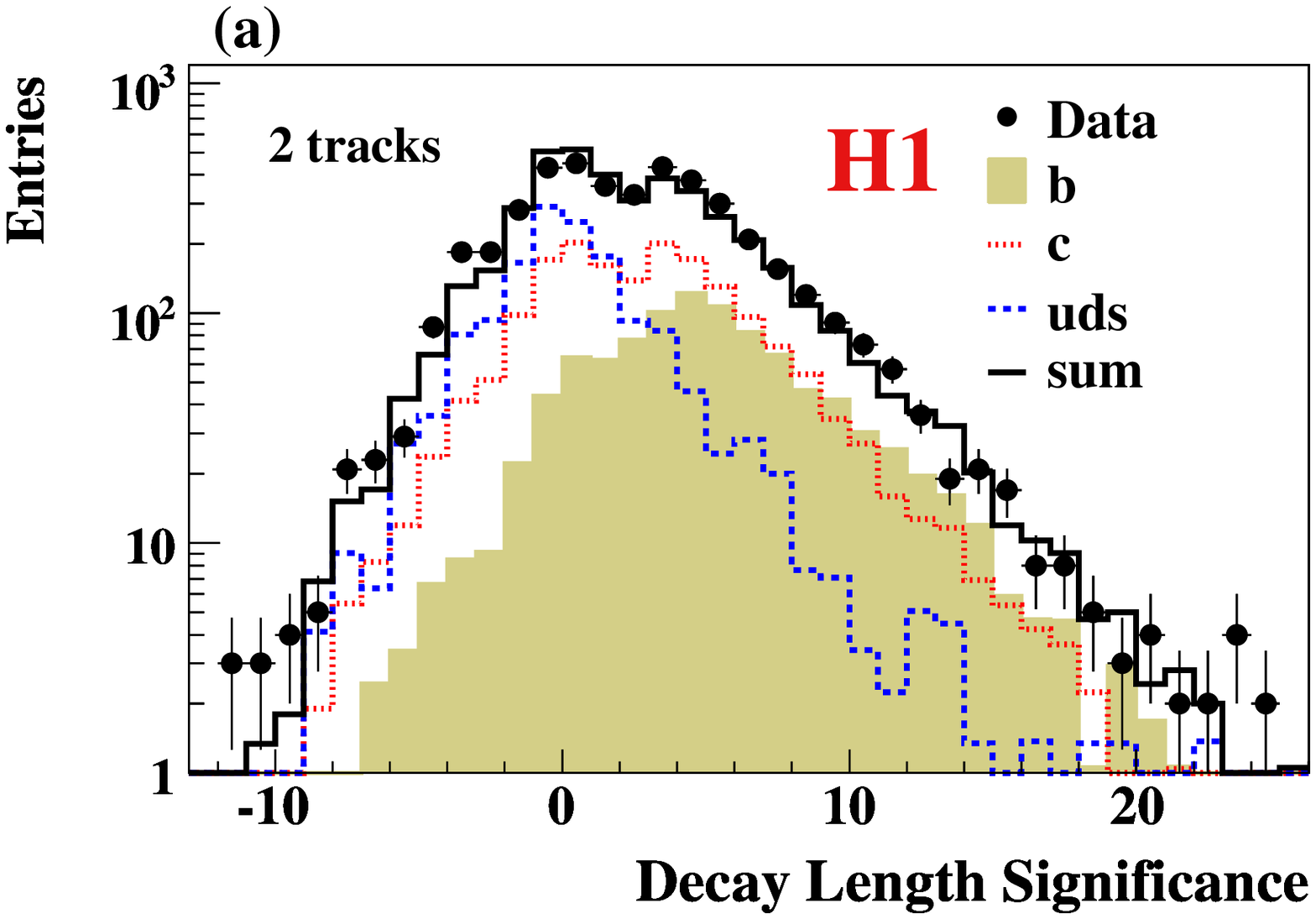}}
\put(3.5,6.){\includegraphics*[width=8.5cm]{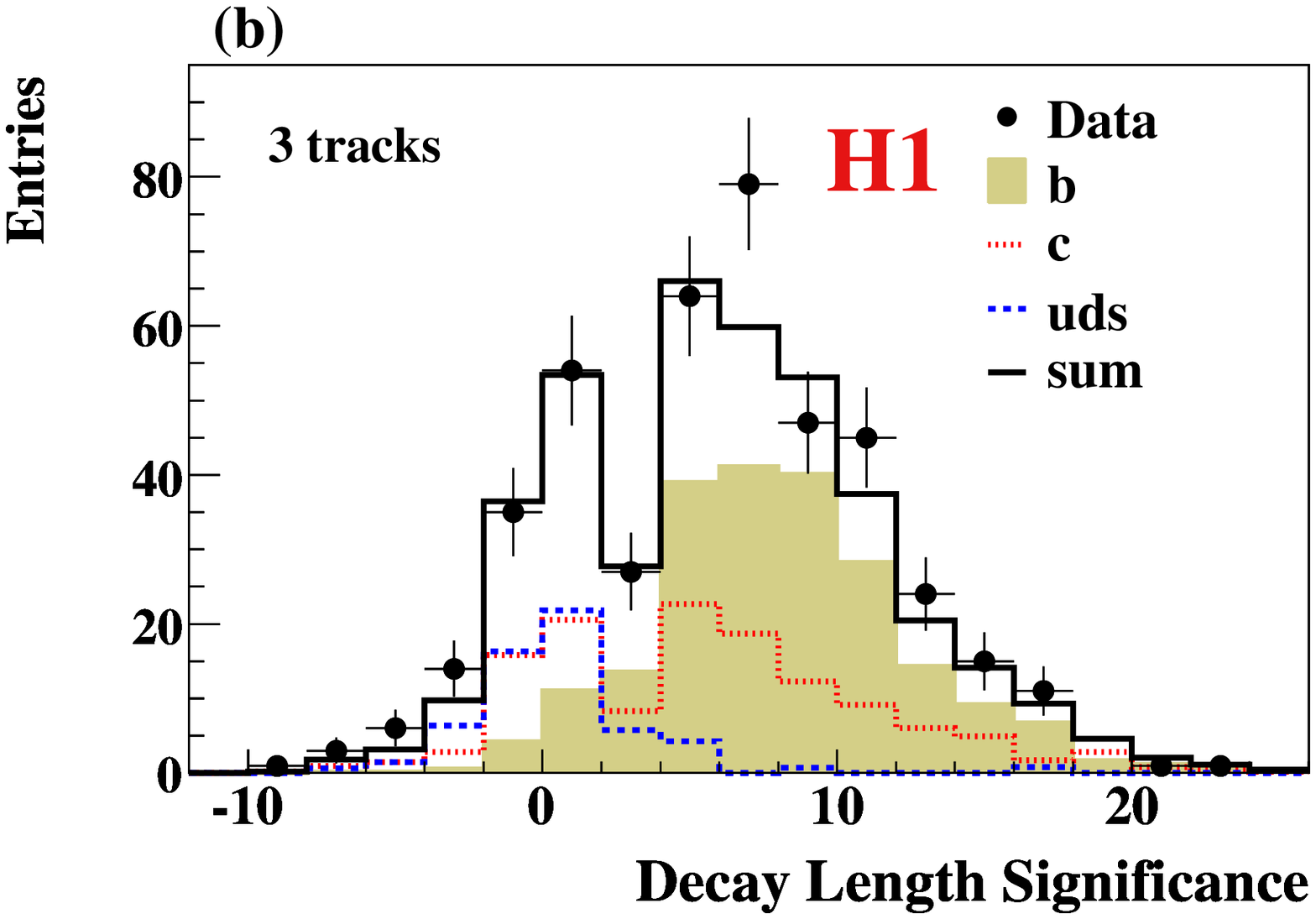}}
\put(3.5,0.){\includegraphics*[width=8.5cm]{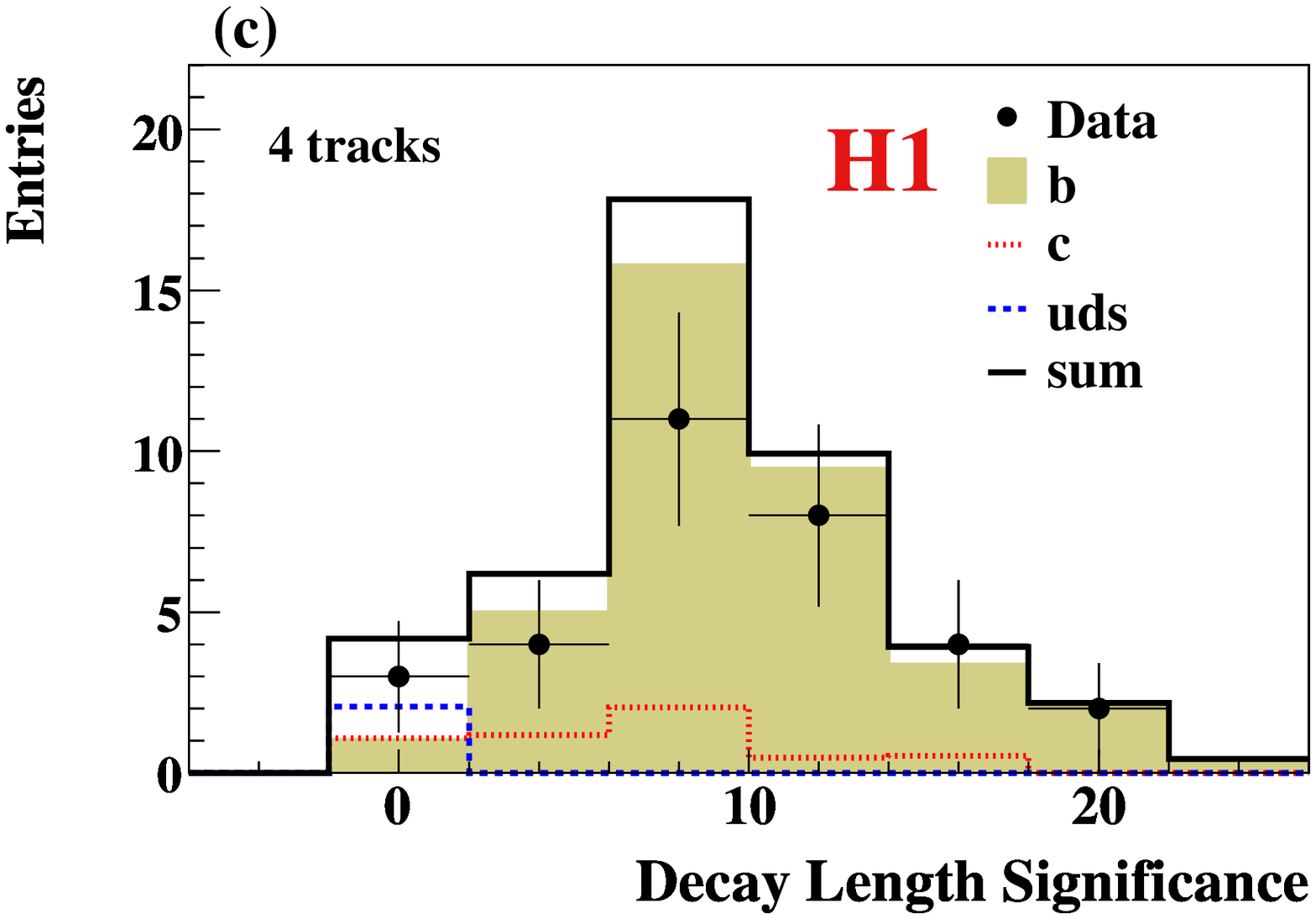}}
\end{picture}
\caption{Decay length significance distributions for samples of events with 
a secondary vertex reconstructed from a) 2, b) 3 and c) 4 tracks. 
The data (points) are compared with the PYTHIA simulation after applying 
the scale factors as obtained from the fit to the subtracted significance 
distributions of the full sample.}
\label{fig:ctrl:annealing} 
\end{figure}

Using a method presented in a previous H1 publication~\cite{Aktas:2004az} 
the primary vertex and the secondary vertices of the heavy hadron decays 
are reconstructed. For each jet the associated tracks are used to reconstruct a
secondary vertex in an iterative procedure. No definite assignment of tracks 
to vertices is made, but each track is assigned a weight in the range 0 to 1 
for each vertex candidate, using the weight function of~\cite{adaptive}. 
The smaller the distance of the track to the vertex candidate, the larger the 
weight. For each jet the vertex configuration that minimises the global 
fit $\chi^2$ is found iteratively using deterministic annealing~\cite{annealing}.
The number of tracks which contribute with a weight greater than 0.8 to the 
secondary vertex is used as a measure of the decay vertex track multiplicity.
The decay length significance is given by the transverse distance 
between the primary and the secondary vertex divided by its error. 
The sign is defined using the projection of the corresponding 
vector on the jet direction.
Figure~\ref{fig:ctrl:annealing} shows the decay length significance distributions
for different decay track multiplicities. Good agreement is seen between 
the data and the distributions from the PYTHIA Monte Carlo simulation which
are scaled by the factors as obtained from the fit to the subtracted significance 
distributions $S_1$ and $S_2$. Conversely, a simultaneous fit of the decay 
length significance distributions for the different track multiplicities gives 
results consistent with those from the subtracted significance distributions.

Figure\,\ref{fig:ctrl:annealing} shows that in heavy quark events secondary 
vertices are significantly displaced in the direction of the related jet axis. 
Contributions from light quarks are mainly observed in the two-track sample at small 
decay length significances. In the two-track sample the contribution from 
charm and beauty is similar to that of light quark events while the three 
and four-track samples are dominated by the beauty component, as expected
from the mean charged particle multiplicity in heavy hadron decays.

\begin{figure}[htb]
\unitlength1cm
\begin{picture}(8,18)
\put(0.,12.){\includegraphics*[width=7.9cm]{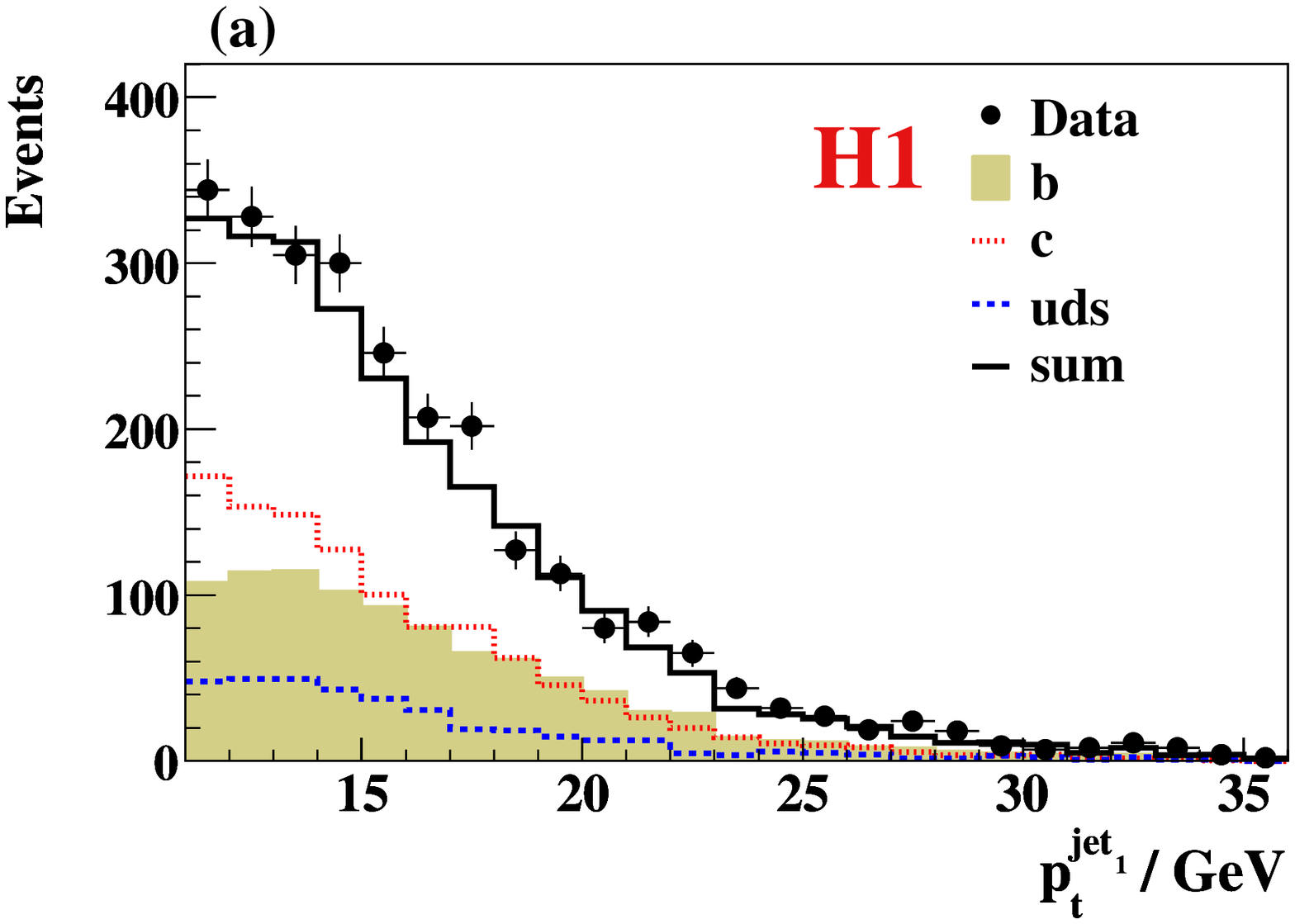}}
\put(8.1,12.){\includegraphics*[width=7.9cm]{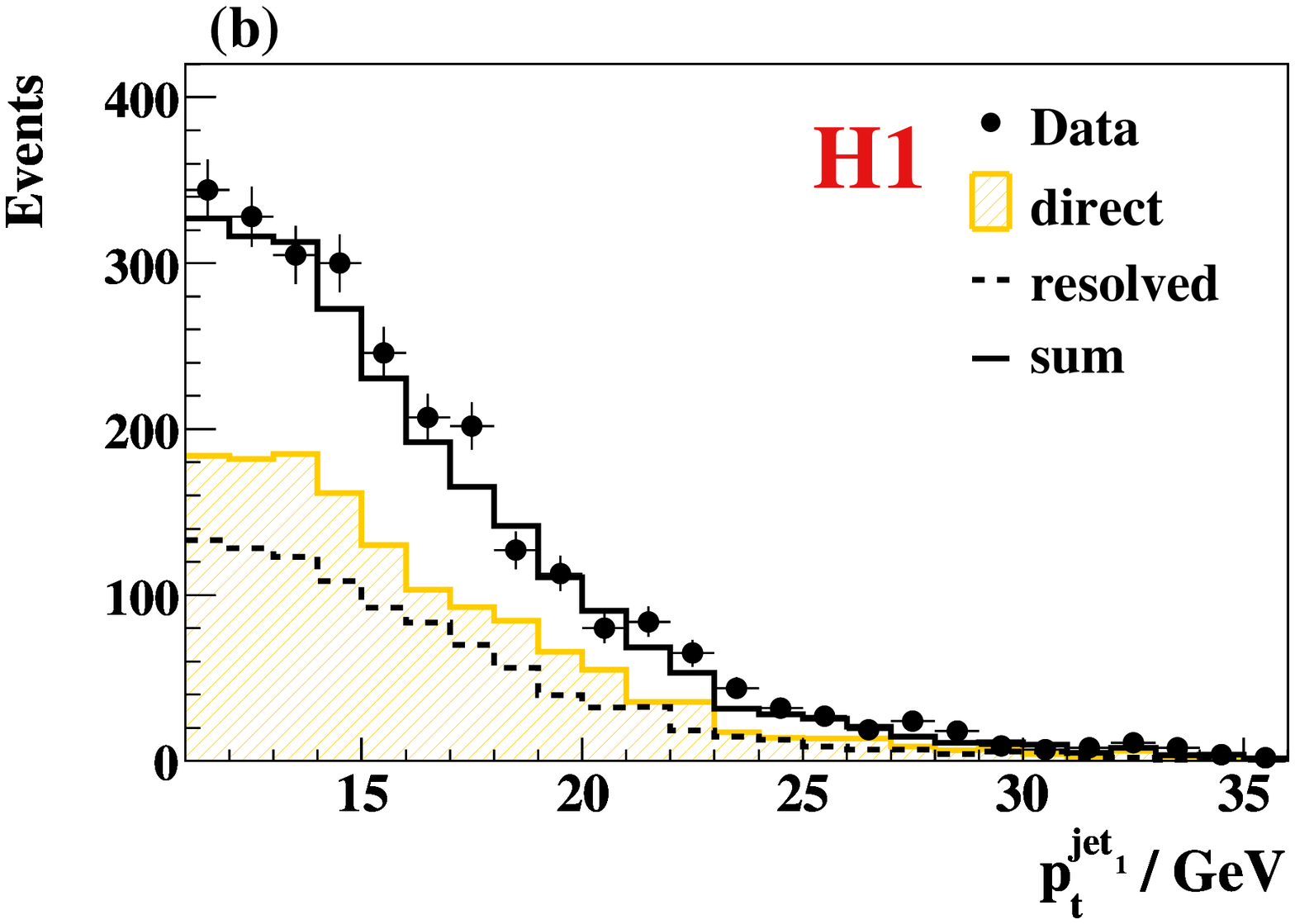}}
\put(0.,6.) {\includegraphics*[width=7.9cm]{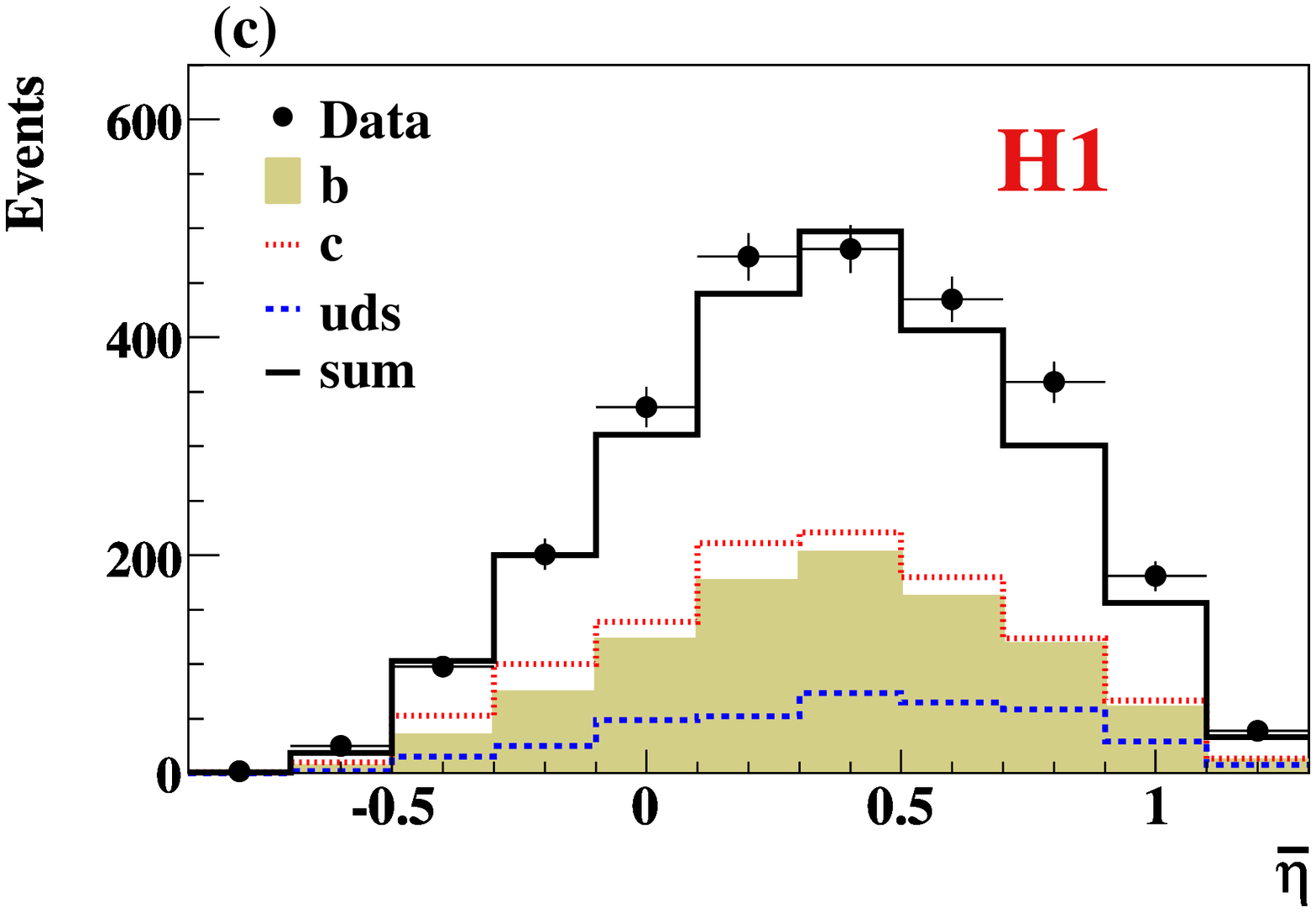}}
\put(8.1,6.){\includegraphics*[width=7.9cm]{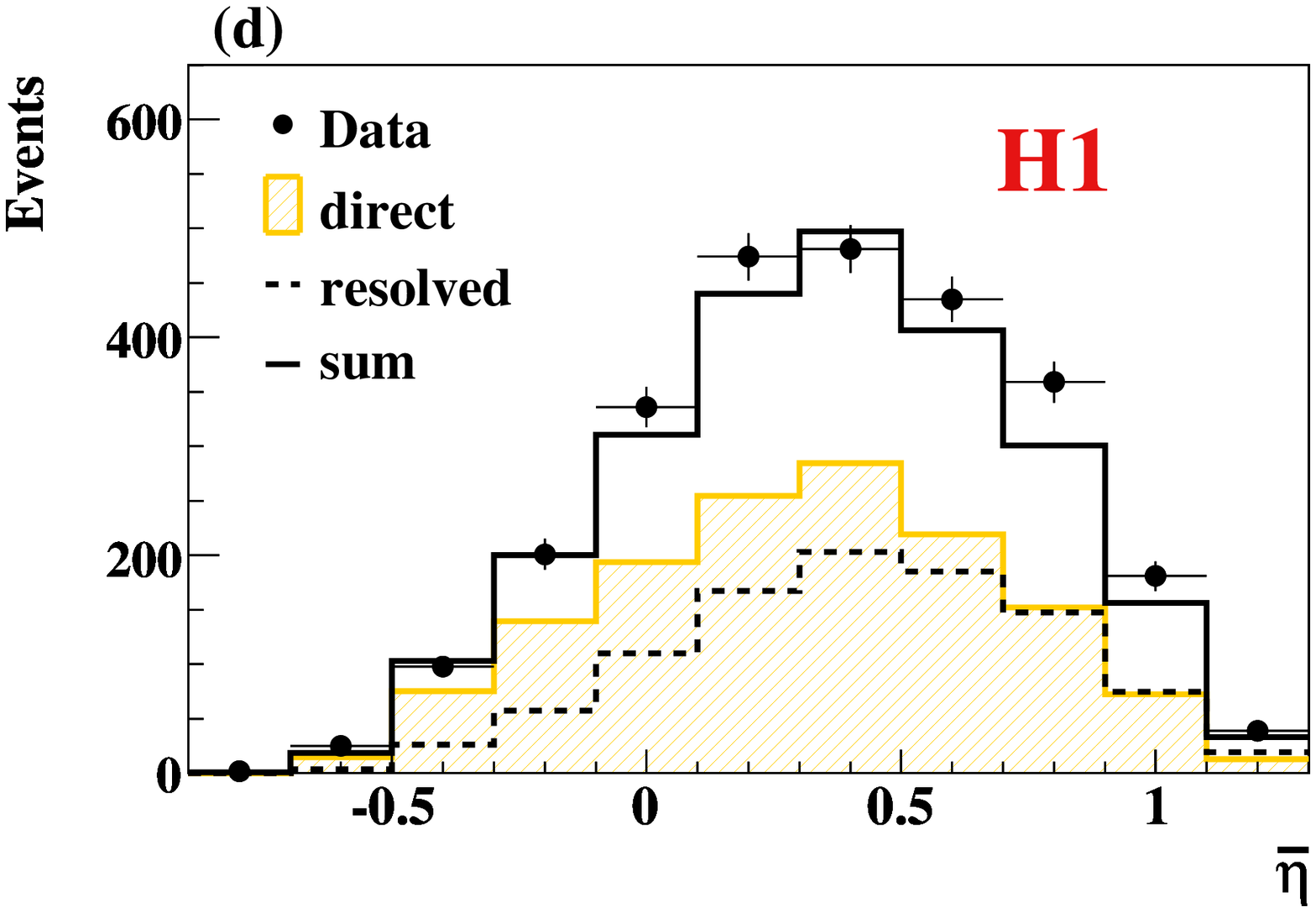}}
\put(0.,0. ){\includegraphics*[width=7.9cm]{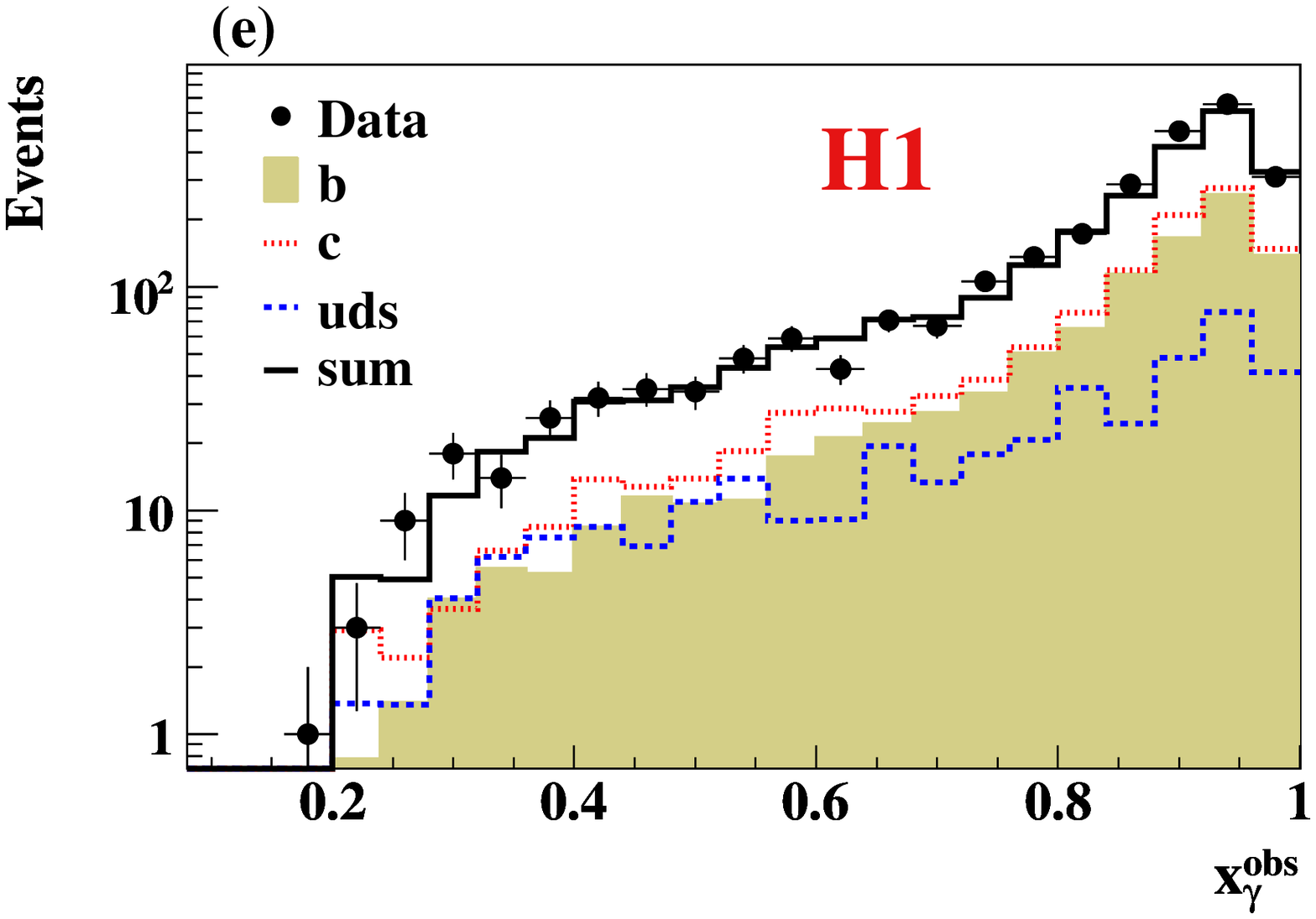}}
\put(8.1,0. ){\includegraphics*[width=7.9cm]{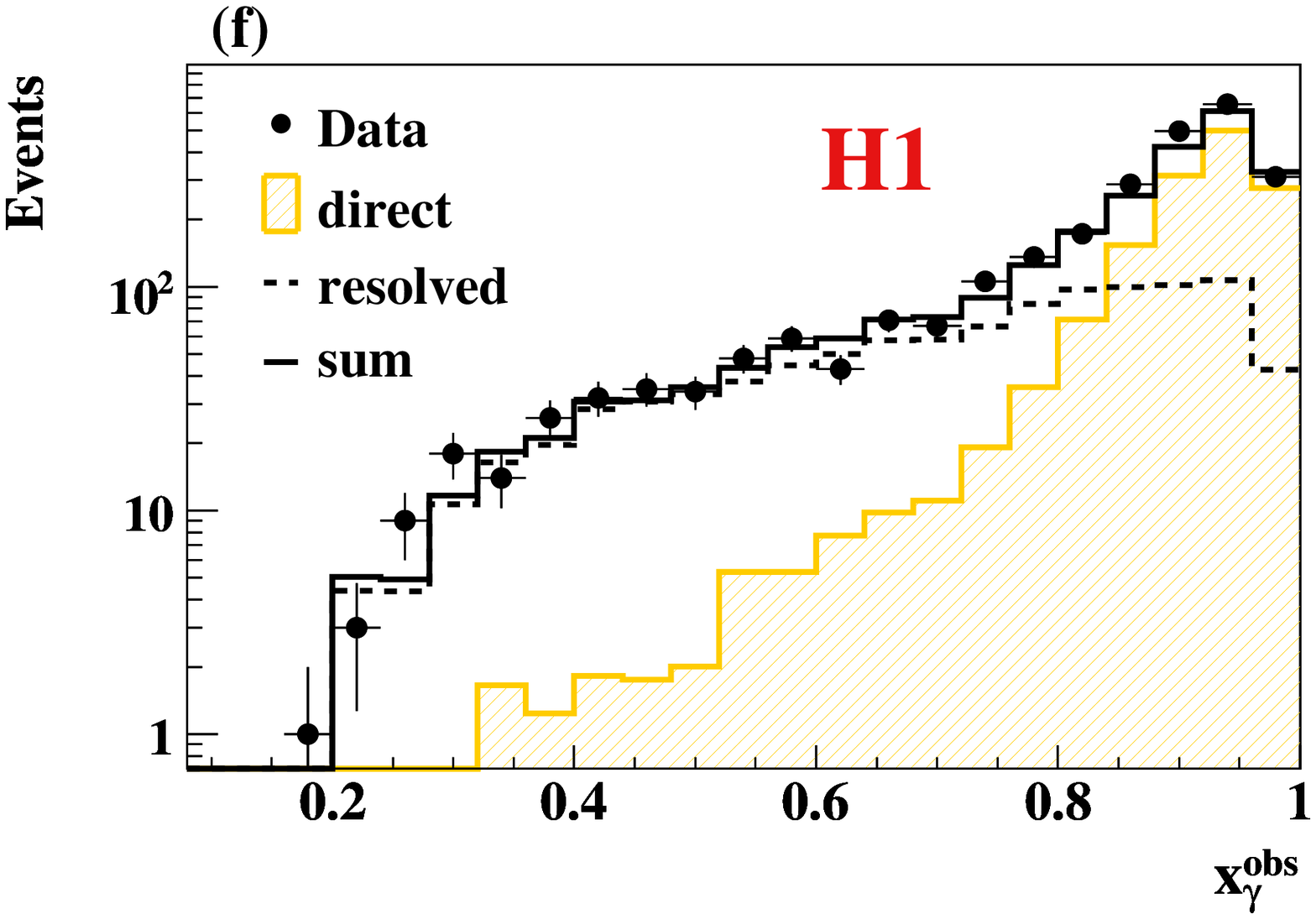}}

\end{picture}
\caption{Distributions of the heavy quark enriched sample with two or more
tracks originating from a secondary vertex and a decay length significance larger than
2.0 (see text).
The data are compared to the PYTHIA simulation for the distributions of
  a,b) the transverse momentum \ptjet \ of the leading jet,
  c,d) the mean pseudo-rapidity $\bar{\eta}$ of the two jets and
  e,f) the observable $\xgobsm$.
The left column shows the decomposition of the distribution into
beauty, charm and light quarks after scaling the PYTHIA predictions
by the scale factors obtained from 
the fit to the subtracted significance distributions of the full sample.
In the right column the contributions from direct and resolved
processes in PYTHIA are indicated as shaded histogram and dashed line 
respectively.}
\label{fig:ctrl:highpurity} 
\end{figure}

The fraction of events with heavy quarks is enhanced
by requiring a secondary decay vertex with two or more tracks and a decay length 
significance larger than 2.0. Applying these cuts, the distributions for \ptjet, 
for $\bar{\eta}$ and for $\xgobsm$ are shown in figure~\ref{fig:ctrl:highpurity}.
The data are compared to the PYTHIA Monte Carlo simulations.
In figures~\ref{fig:ctrl:highpurity}a, c and e the decomposition into light, 
charm and beauty quark contributions is indicated.
These are determined using the scale factors as obtained from the fit to the 
subtracted significance distributions of the full sample. Good agreement between 
the PYTHIA prediction and the data is seen. The contributions from charm and beauty 
events are $44\%$ and $41\%$ respectively, while $15\%$ of the events contain only 
light quarks.
In figures~\ref{fig:ctrl:highpurity}b, d and f the 
same data are shown together with the contributions from direct 
($\gamma g \rightarrow q\bar{q}$) and resolved processes, as predicted
by the PYTHIA simulation. According to the leading order QCD calculation, as 
implemented in PYTHIA, the fraction of events that arise from processes
with resolved photons is about $40\%$.

In the region $\xgobsm<0.85$ the contribution from resolved
processes is enhanced and amounts to about $80\%$. According to PYTHIA
the contribution from processes with heavy quark excitation is by
far dominant. 
%
%
In conclusion, the data are well described by the leading order Monte Carlo 
simulation PYTHIA in which significant contributions from heavy quark 
excitation processes are predicted.

\section{Conclusions}
\label{sec:conclusions}

Differential charm and beauty dijet photoproduction cross 
sections at HERA are measured using a technique based on the
lifetime of the heavy quark hadrons.
The heavy quark cross sections are determined using the subtracted
impact parameter significance distributions of tracks in dijet events.
The cross sections are measured as functions of the transverse momentum
\ptjet \ of the leading jet, of the mean pseudo-rapidity $\bar{\eta}$ 
of the two jets and of the observable \xgobs. 
Taking into account the theoretical uncertainties,
the charm cross sections are consistent both in normalisation and shape
with a calculation in perturbative QCD to next-to-leading order.
The beauty cross sections tend to be higher than NLO, 
by $1.6 \sigma$ for the total cross section, 
with a more significant discrepancy observed in the region of $\xgobsm<0.85$ 
where processes involving resolved photons or higher order contributions
are expected to be enhanced. 
The Monte Carlo generator PYTHIA gives a good description of the shape 
of both the charm and the beauty data.
The data confirm the expectation within this model that a significant contribution 
to the heavy quark dijet cross section comes from processes with resolved photons. 
In the region $\xgobsm>0.85$, the relative contributions from charm and 
beauty to the inclusive dijet cross section are found to be in agreement 
within errors with values of 4/11 and 1/11, i.e.\,the na\"ive expectation for the 
direct photon-gluon fusion process, assuming all quarks to be massless.

\section*{Acknowledgements}

We are grateful to the HERA machine group whose outstanding
efforts have made this experiment possible.
We thank the engineers and technicians for their work in constructing and
maintaining the H1 detector, our funding agencies for
financial support, the DESY technical staff for continual assistance
and the DESY directorate for support and for the
hospitality which they extend to the non-DESY
members of the collaboration.

\clearpage


\begin{table}[h]
\begin{center}
\begin{tabular}{|rr|r||rrr|rrr|}
\hline 
 \multicolumn{3}{|c||}{} &
 \multicolumn{6}{|c|}{Charm} \\ 
         \hline \hline
         \multicolumn{2}{|c|}{\rule[-2mm]{0mm}{7mm} \ptjet \ range} & $\langle \ptjetm \rangle$ & 
         $d\sigma/d\ptjetm$ & stat. & syst. & \fccb \ & stat. & syst. \\
 \multicolumn{2}{|c|}{[GeV]} & [GeV] & \multicolumn{3}{c|}{[pb/GeV]} & \multicolumn{3}{c|}{} \\ 
\hline
 11.0  & 14.5  & 12.75  & 137  & 20  & 20 & 0.333 & 0.048 & 0.044 \\
 14.5  & 18.0  & 16.25  & 49.5   & 7.6   & 7.9  & 0.352 & 0.054 & 0.046 \\
 18.0  & 22.5  & 20.0   & 10.6   & 2.5   & 1.8  & 0.270 & 0.062 & 0.035 \\
 22.5  & 35.0  & 27.0   & 2.46   & 0.86  & 0.45 & 0.402 & 0.140 & 0.052 \\
\hline \hline
         \multicolumn{2}{|c|}{\rule[-2mm]{0mm}{7mm} $\bar{\eta}$ range} & $\langle\bar{\eta}\rangle$ & 
         $d\sigma/d\bar{\eta}$ & stat. & syst. & \fccb \ & stat. & syst. \\
 \multicolumn{2}{|c|}{} &  & \multicolumn{3}{c|}{[pb]} & \multicolumn{3}{c|}{} \\ 
\hline

 $-$0.90  &  0.10 & $-$0.35  & 162   & 24  & 22  & 0.337 & 0.051 & 0.045 \\
  0.10  &  0.60 &  0.35      & 674   & 86  & 101 & 0.346 & 0.044 & 0.045 \\
  0.60  &  1.30 &  0.95      & 386   & 52  & 61  & 0.358 & 0.049 & 0.047 \\
 
          \hline \hline
         \multicolumn{2}{|c|}{\rule[-2mm]{0mm}{7mm} $\xgobsm$ range} & $\langle\xgobsm\rangle$ & 
         $d\sigma/d\xgobsm$ & stat. & syst. & \fccb \ & stat. & syst. \\
 \multicolumn{2}{|c|}{} &  & \multicolumn{3}{c|}{[pb]} & \multicolumn{3}{c|}{} \\ 
\hline
  0.10  & 0.70  & 0.45  & 236  & 70   & 34  & 0.250 & 0.074 & 0.033 \\  
  0.70  & 0.85  & 0.77  & 1017 & 271  & 162 & 0.342 & 0.091 & 0.047 \\  
  0.85  & 1.00  & 0.92  & 2994 & 280  & 498 & 0.374 & 0.035 & 0.049 \\  
\hline 
 \multicolumn{9}{c}{} \\
\hline 
 \multicolumn{3}{|c||}{} &
 \multicolumn{6}{|c|}{Beauty} \\
         \hline \hline
         \multicolumn{2}{|c|}{\rule[-2mm]{0mm}{7mm} \ptjet \ range} & $\langle \ptjetm \rangle$ & 
         $d\sigma/d\ptjetm$ & stat. & syst. & \fbbb \ & stat. & syst. \\
 \multicolumn{2}{|c|}{[GeV]} & [GeV] & \multicolumn{3}{c|}{[pb/GeV]} & \multicolumn{3}{c|}{} \\ 
\hline
   11.0  & 14.5  & 12.75  &  24.7  & 6.1  & 5.5  & 0.060 & 0.015 & 0.013\\
   14.5  & 18.0  & 16.25  &  9.79  & 1.87  & 2.15  & 0.070 & 0.013 & 0.015\\
   18.0  & 22.5  & 20.0  &   3.37  & 0.48 & 0.78 & 0.085 & 0.012 & 0.018\\
   22.5  & 35.0  & 27.0  &   0.28  & 0.14 & 0.07 & 0.046 & 0.022 & 0.010\\
          \hline \hline
         \multicolumn{2}{|c|}{\rule[-2mm]{0mm}{7mm} $\bar{\eta}$ range} & $\langle\bar{\eta}\rangle$ & 
         $d\sigma/d\bar{\eta}$ & stat. & syst. & \fbbb \ & stat. & syst. \\
 \multicolumn{2}{|c|}{} &  & \multicolumn{3}{c|}{[pb]} & \multicolumn{3}{c|}{} \\ 
\hline

  $-$0.90  &  0.10 & $-$0.35  &  37.1  & 5.9  & 7.8  & 0.077 & 0.012 & 0.016\\
   0.10  &  0.60 &  0.35  &  112 & 17 & 25 & 0.057 & 0.009 & 0.012\\
   0.60  &  1.30 &  0.95  &  56.2  & 11.3 & 12.4 & 0.052 & 0.010 & 0.011\\
          \hline \hline
         \multicolumn{2}{|c|}{\rule[-2mm]{0mm}{7mm} $\xgobsm$ range} & $\langle\xgobsm\rangle$ & 
         $d\sigma/d\xgobsm$ & stat. & syst. & \fbbb \ & stat. & syst. \\
 \multicolumn{2}{|c|}{} &  & \multicolumn{3}{c|}{[pb]} & \multicolumn{3}{c|}{} \\ 
\hline
  0.10  & 0.70  & 0.45  &  44.6  & 12.8 & 9.9   & 0.047 & 0.014 & 0.010\\
  0.70  & 0.85  & 0.77  &  191 & 60 & 44  & 0.064 & 0.020 & 0.014\\
  0.85  & 1.00  & 0.92  &  584 & 72 & 135 & 0.073 & 0.009 & 0.015\\
\hline
\end{tabular}
\caption{The measured charm and beauty dijet photoproduction cross sections 
and the relative contributions \fccb \ and \fbbb \ 
to the inclusive dijet photoproduction cross section
with statistical and systematic errors.}

\label{tab:xsecs:inclusive}
\end{center}
\end{table}

\begin{table}[h]
\begin{center}
\begin{tabular}{|rr|r||rrr|rrr|}
\hline 
 \multicolumn{3}{|c||}{} &
 \multicolumn{6}{|c|}{Charm} \\ 
         \hline \hline
         \multicolumn{2}{|c|}{\rule[-2mm]{0mm}{7mm} \ptjet \ range} & $\langle \ptjetm \rangle$ & 
         $d\sigma/d\ptjetm$ & stat. & syst. & \fccb \ & stat. & syst. \\
 \multicolumn{2}{|c|}{[GeV]} & [GeV] & \multicolumn{3}{c|}{[pb/GeV]} & \multicolumn{3}{c|}{} \\ 
\hline
 11.0  & 14.5  & 12.75  & 73.3  & 12.7  & 10.7 & 0.348 & 0.060 & 0.046 \\
 14.5  & 18.0  & 16.25  & 29.0  &  4.6  & 4.6  & 0.364 & 0.057 & 0.048 \\
 18.0  & 22.5  & 20.0   & 8.79  &  1.75  & 1.46  & 0.357 & 0.071 & 0.047 \\
 22.5  & 35.0  & 27.0   & 1.94  &  0.70 & 0.35 & 0.455 & 0.165 & 0.060 \\
          \hline \hline
         \multicolumn{2}{|c|}{\rule[-2mm]{0mm}{7mm} $\bar{\eta}$ range} & $\langle\bar{\eta}\rangle$ & 
         $d\sigma/d\bar{\eta}$ & stat. & syst. & \fccb \ & stat. & syst. \\
 \multicolumn{2}{|c|}{} &  & \multicolumn{3}{c|}{[pb]} & \multicolumn{3}{c|}{} \\ 
\hline

 $-$0.90  &  0.10 & $-$0.35  & 115  & 21  & 16 & 0.322 & 0.058 & 0.043 \\
  0.10  &  0.60 &  0.35      & 496  & 70  & 74 & 0.462 & 0.065 & 0.060 \\
  0.60  &  1.30 &  0.95      & 165  & 24  & 26 & 0.374 & 0.055 & 0.049 \\
\hline 
 \multicolumn{9}{c}{} \\
\hline 
 \multicolumn{3}{|c||}{} &
 \multicolumn{6}{|c|}{Beauty} \\
         \hline \hline
         \multicolumn{2}{|c|}{\rule[-2mm]{0mm}{7mm} \ptjet \ range} & $\langle \ptjetm \rangle$ & 
         $d\sigma/d\ptjetm$ & stat. & syst. & \fbbb \ & stat. & syst. \\
 \multicolumn{2}{|c|}{[GeV]} & [GeV] & \multicolumn{3}{c|}{[pb/GeV]} & \multicolumn{3}{c|}{} \\ 
\hline 
   11.0  & 14.5  & 12.75  &   15.3  & 4.3  & 3.4  & 0.073 & 0.020 & 0.015\\
   14.5  & 18.0  & 16.25  &   6.19  & 1.20 & 1.42 & 0.078 & 0.015 & 0.016\\
   18.0  & 22.5  & 20.0   &   1.89  & 0.31 & 0.44 & 0.076 & 0.013 & 0.016\\
   22.5  & 35.0  & 27.0   &   0.27  & 0.18 & 0.07 & 0.063 & 0.043 & 0.014\\
          \hline \hline
         \multicolumn{2}{|c|}{\rule[-2mm]{0mm}{7mm} $\bar{\eta}$ range} & $\langle\bar{\eta}\rangle$ & 
         $d\sigma/d\bar{\eta}$ & stat. & syst. & \fbbb \ & stat. & syst. \\
 \multicolumn{2}{|c|}{} &  & \multicolumn{3}{c|}{[pb]} & \multicolumn{3}{c|}{} \\ 
\hline

  $-$0.90  &  0.10 & $-$0.35  &  29.8   & 4.7  & 6.3  & 0.084 & 0.013 & 0.018 \\
   0.10  &  0.60 &  0.35  &  54.4   & 15.5 & 12.0 & 0.051 & 0.014 & 0.011 \\
   0.60  &  1.30 &  0.95  &  31.9   & 5.7  & 7.1  & 0.072 & 0.013 & 0.016 \\
\hline

\end{tabular}
\caption{The measured charm and beauty dijet photoproduction cross sections and 
the relative contributions \fccb \ and \fbbb \ to the inclusive photoproduction 
dijet cross section for the region $\xgobsm > 0.85$ with statistical and systematic errors.}
\label{tab:xsecs:direct}
\end{center}
\end{table}

\newpage



\end{document}